\documentclass[11pt]{article}
\pdfoutput=1
\usepackage{graphicx}
\usepackage{hyperref}
\usepackage{amsmath}
\usepackage{amssymb}
\usepackage[T1]{fontenc}
\usepackage[font=small,labelfont=bf]{caption}
\usepackage{cite}

\newcommand{\beq}{\begin{eqnarray}}
\newcommand{\eeq}{\end{eqnarray}}
\def\simlt{\stackrel{<}{{}_\sim}}
\def\simgt{\stackrel{>}{{}_\sim}}

\begin{document}

\title{{\bf \LARGE Double-Disk Dark Matter}}

\author{{\Large JiJi Fan, Andrey Katz, Lisa Randall, and Matthew Reece}}
\date{}

\maketitle
\vspace{-1.1cm}
\begin{center}
{\it Department of Physics, Harvard University, Cambridge, MA, 02138}
\end{center}

\begin{abstract}
\noindent Based on observational tests of large scale structure and constraints on halo structure, dark matter is generally taken to be cold and essentially collisionless.
On the other hand, given the large number of particles and forces in the visible world,  a more complex dark sector could be a reasonable or even likely possibility.
 This hypothesis leads to testable consequences, perhaps portending the  discovery of a rich hidden world neighboring our own. We consider a scenario that readily satisfies
current bounds that we call
 Partially Interacting Dark Matter (PIDM). This scenario contains self-interacting dark matter, but it is not the dominant
 component. Even if PIDM contains only a fraction of the net dark matter density, comparable to the baryonic fraction, the subdominant component's interactions
can lead to interesting and potentially observable consequences. Our primary focus will be the special case of Double-Disk Dark Matter (DDDM), in which  self-interactions
allow the dark matter to lose enough energy to lead to dynamics similar to those in the baryonic sector. We explore a simple model in which DDDM can cool efficiently and
form a disk within galaxies, and we evaluate some of the possible observational signatures. The most prominent signal of such a scenario could be an enhanced indirect
detection signature with a distinctive spatial distribution. Even though subdominant, the enhanced density at the center of the galaxy and possibly throughout the plane
of the galaxy (depending on precise alignment) can lead to large boost factors, and could even explain a signature as large  as the 130 GeV Fermi line. Such scenarios
also predict additional dark radiation degrees of freedom that could soon be detectable and would influence the interpretation of future data, such as that from Planck and
from the Gaia satellite. We consider this to be the first step toward exploring a rich array of new
possibilities for dark matter dynamics.

\end{abstract}

\section{Introduction}
%%%%%%%%%%%%%%%%%%%%%%%%%%%%%%%%%%%%%%%%

All known particles make up only a small fraction of the energy density in our universe, yet the Standard Model is extremely complicated: three forces, one Higgsed, one confining, plus quarks and leptons organized into three generations. This model---the components of the visible universe---deviates markedly from any apparent principle of minimality. Yet, when considering the 85\% of the matter in the universe that is dark, our usual response is to turn to minimal models of a cold, collisionless particle: a WIMP, perhaps, or an axion. Slightly less minimal variations are sometimes studied, often motivated by data that is in mild conflict with the cold dark matter paradigm. Self-interacting dark matter (SIDM)~\cite{Spergel:1999mh} and warm dark matter~\cite{Bode:2000gq} are well-studied examples. These minimal choices are, to some extent, justified by Ockham's razor. We know that the bulk of galaxy halos consists of dark matter organized into large, diffuse, spheroidal distributions, and (based on halo shapes and the Bullet Cluster) that the stuff making up these halos is approximately collisionless, which offers some support for the idea of minimality. Still, confronted with the richness of physics in the visible world around us, it is tantalizing to imagine that the dark world could be similarly complex, full of structures, forces, and matter that are invisible to us. We might hope that a whole sector of the universe as rich as our own exists just out of sight. Our goal in this paper is to argue that this superficially fanciful idea should be taken seriously as a testable hypothesis, which might even help to resolve some of the deficiencies of the $\Lambda$CDM scenario. Double-Disk Dark Matter (DDDM) is a concrete form of this idea, in which a small fraction of all dark matter has dissipative dynamics causing it to cool into a disk within the Milky Way galaxy. Double-Disk Dark Matter acts less like typical noninteracting dark matter than like a new kind of ordinary matter, constituting an invisible world that may be literally parallel to our own.

The general scenario that we propose, Partially Interacting Dark Matter (PIDM), is that a subdominant component of dark matter has self-interactions. The more specific DDDM scenario that we focus on assumes the existence of a massless (or nearly massless) U(1) gauge boson that permits dissipative dynamics. This will generate observationally distinguishable dark matter consequences and in particular a thin dark matter disk similar to the baryonic disk when an additional light dark charged particle is present with sufficient abundance. Even though by assumption the dark matter is subdominant, the density of the interacting component in the disk could be much higher than the dominant diffuse dark matter that is  spread throughout the halo and could hence lead to stronger indirect signatures. These could include observations such as the recently reported Fermi line~\cite{Bringmann:2012vr,Weniger:2012tx}, which is challenging to explain without significant boost factors or tuned scenarios (many of which are summarized in~\cite{Buckley:2012ws,Tulin:2012uq}). We will construct models that generate this signal to illustrate one possible consequence of DDDM, but the potential signatures  span a much wider range of possibilities. Data will determine whether a richer dark matter sector does in fact exist.

 Because existing constraints are weak as we will see in Section~\ref{sec:pddmamount}, the fraction of additional interacting dark matter is fairly unconstrained so long as it is subdominant and nondissipative. When the dark matter does collapse into a disk, the constraint is stronger due to the Oort limit, which is a bound on the amount of matter in the vicinity of our solar system. For many of the numerical results in this paper, we will assume that the energy fraction is at the Oort limit, yielding a fraction of the energy density comparable to but smaller than that of baryons.   Because new forms of matter might be present at the weak scale, and because our scenario was motivated in part by the Fermi photon line, we will often assume a new interacting dark matter particle of mass near 100 GeV. Again, the scenario allows for a wide range of parameters and we consider this mass simply for specificity.

Partially Interacting Dark Matter and the Double-Disk Dark Matter scenario that we focus on can potentially lead to other interesting consequences. In future work we would like to explore the effect of this component on overall structure formation and on observations such as gravitational lensing and detailed sky surveys. This scenario can also lead to an acoustic oscillation signature similar to that from baryons~\cite{CyrRacine:2012fz}. Detailed observations of the CMBR and galaxy correlation functions should also detect or constrain DDDM.

 We note that other authors have proposed interesting scenarios for dark matter involving long range forces and bound states: the older idea of mirror matter~\cite{Khlopov:1989fj,Berezhiani:1995am,Mohapatra:2000qx,Mohapatra:2001sx,Foot:2004pa,Silagadze:2008fa,Higaki:2013vuv}, recent work on dark matter charged under a hidden U(1)~\cite{Feng:2008mu,Ackerman:2008gi,Feng:2009mn}, and more recent work on dark atoms~\cite{Kaplan:2009de,Kaplan:2011yj,Cline:2012is,CyrRacine:2012fz}. In fact, all of these ideas were foreshadowed in much earlier work by Goldberg and Hall~\cite{Goldberg:1986nk}. A few other scenarios including DM without conserved particle number~\cite{Carlson:1992fn}, dynamical dark matter~\cite{Dienes:2011ja,Dienes:2011sa}, or DM interacting through long-range scalar forces~\cite{Baldi:2012kt,Baldi:2012ua} involve very different physics but the same spirit of exploring nontrivial dark sector dynamics. Furthermore, the idea of self-interacting dark matter (SIDM) has been studied intensively as a possible solution to discrepancies between cold dark matter models and observations (e.g. the cusp/core problem~\cite{Flores:1994gz,Moore:1994yx}), beginning with Ref.~\cite{Spergel:1999mh,Dave:2000ar} (for recent progress see Refs.~\cite{Rocha:2012jg,Peter:2012jh}). These scenarios are often very constrained by observations of the halo and galaxy and galaxy cluster interactions. The chief difference in our scenario is that the interacting component of dark matter is subdominant. We will see that such a scenario is far less restrictive since galaxy shape constraints as well as direct constraints on interactions rely predominantly on the existence of a dominant noninteracting component. A subdominant component can interact and permit much richer dissipative dynamics and that is what we consider below.

We begin by explaining in Section~\ref{sec:pddmamount} how current constraints can allow interesting amounts of interacting dark matter. In Section~\ref{sec:history} we explain that they also allow background dark radiation in the amount predicted by our model, which will be probed in the near future by the Planck satellite. In section~\ref{sec:fermiline} we argue, using the tentative Fermi line as an example, that possible gamma ray line signals of dark matter annihilation at the level probed by current observations are difficult to explain without a very large boost factor. In section~\ref{sec:cooling}, we show that it is possible for a subdominant species of dark matter to efficiently cool within the lifetime of the universe. We follow this with a discussion of the structure that forms as a result of cooling, the DDDM disk, in section~\ref{sec:diskform}. At this point, having established that a disk can form and estimated how much matter can lie in it, we are equipped to return to the issue of a boost factor for indirect detection, which we discuss (along with direct detection prospects) in section~\ref{sec:detection}. Even if it does not provide such signatures, DDDM could exist and be detected through its gravitational effects. We conclude in section~\ref{sec:conclusions} with an outline of the many exciting possibilities for future work on this subject.

%%%%%%%%%%%%%%%%%%%%%%
\section{Constraining the Amount of Allowed DDDM Abundance}
\label{sec:pddmamount}
%%%%%%%%%%%%%%%%%%%%%%%%%%%%%%%%%%%%%%%%

We define the fraction of the energy density in PIDM, compared to ordinary dark matter, as $\epsilon_\Omega \equiv \frac{\Omega_{\rm PIDM}}{\Omega_{\rm DM}}$. Furthermore, we expect the relative fraction of different matter components in the Milky Way is comparable to that in the universe as a whole.  So  we  take $\epsilon_\Omega \approx M_{\rm PIDM}^{\rm gal}/M_{\rm DM}^{\rm gal}$ where $M_{\rm PIDM}^{\rm gal}$ is the total mass of PIDM in the galaxy and $M_{\rm DM}^{\rm gal}$ is the total mass of all dark matter in the galaxy. In fact, in all likelihood the ratio can be bigger in terms of the total energy accounting, since our strongest bound is only on PIDM in disk form and only about a third of the baryons end up in the disk. For this case of DDDM, we denote the fraction of mass in the Milky Way's disk by
\beq
\epsilon \equiv \frac{M^{\rm disk}_{\rm DDDM}}{M^{\rm gal}_{\rm DM}}.
\eeq
If DDDM is organized similarly to baryons, the total energy fraction in DDDM would be more like $\epsilon_\Omega \approx 3 \epsilon$. But in most of the paper we will take $\epsilon_\Omega \approx \epsilon$ for simplicity.

Current bounds on self-interacting dark matter arise from halo shapes and cluster interactions. So far such bounds have been calculated only for a single dark matter component, for which they can be quite constraining. Self-interactions lead to more spherical halos, especially in the inner region, where the density is higher and interactions are more frequent~\cite{Dave:2000ar,Rocha:2012jg}.

One bound of this type arises from the halo for the galaxy cluster MS 2137-23~\cite{MiraldaEscude:2000qt}, which is measured by gravitational lensing to show a 20\% deviation from axial symmetry at radius 70 kpc from the center. Another bound comes from measurements of X-rays emitted by hot gas in the elliptical galaxy NGC 720~\cite{Buote:2002wd}, showing 35\% deviations from sphericity at distances of 5 to 10 kpc from the center. Roughly speaking, these bounds on SIDM exclude the possibility that a typical dark matter particle has scattered at least once in the age of the universe. The time scale for a dark matter particle to scatter is $\left<n \sigma v\right>^{-1}$, with $n = \rho/m$ the number density of other dark matter particles it could interact with. Hence, the limits are typically expressed as a bound on cross section per unit mass, $\sigma/m$. The inferred bound is approximately $\sigma/m \simlt 0.1 {\rm cm}^2/{\rm g}  \approx 0.2~{\rm barn}/{\rm GeV}$~\cite{Peter:2012jh}. Such cross sections are large by the standards of pointlike particles, but readily arise through long-range forces (for instance, Rutherford scattering has a $1/v^4$ enhancement at low velocities, leading to large cross sections) or through large composite objects (e.g., atoms with Bohr radii $a_0 \sim m^{-1}/\alpha$ at weak coupling $\alpha$, which have cross sections even larger than $4\pi a_0^2$~\cite{Kaplan:2009de}). In the case that all dark matter consists of a particle (and its antiparticle) charged under a massless U(1) gauge boson, these bounds have been studied in Refs.~\cite{Ackerman:2008gi} and~\cite{Feng:2009mn}, which find that because of the low-velocity enhancement of Rutherford scattering the bounds exclude a thermal relic abundance. In the case of composite dark atoms, the bounds have been studied in Refs.~\cite{Kaplan:2009de,CyrRacine:2012fz}. They exclude a portion of the parameter space, but less than in the fully ionized case, because atom/atom scattering is closer to a hard-sphere interaction without a long-range force.

Such bounds do not directly apply to PIDM since a sufficiently small fraction of all matter could have extremely strong interactions without affecting observations at all. A conservative estimate of the allowed abundance of PIDM can be found using the recent halo shape analysis of Ref.~\cite{Peter:2012jh}, which corrects certain deficiencies in earlier analyses and argues that $\sigma/m = 1~{\rm cm}^2/{\rm g}$ is ruled out by the X-ray observations of NGC 720 but that $\sigma/m = 0.1~{\rm cm}^2/{\rm g}$ is allowed by all current bounds. Figure 5 of Ref.~\cite{Peter:2012jh} shows that 20\% to 30\% deviations from spherical symmetry, as observed in data, are compatible with all dark matter particles scattering a few times in the age of the universe. In that case, 10\% of all the DM would have scattered 10 times or more. PIDM---which will scatter multiple times and form denser structures---together with 90\% ordinary dark matter with no interactions at all, will leave the halo triaxial. Numerical simulations should be done to quantify these statements more carefully and produce a definite bound, but we summarize by saying that halo shapes are compatible with  10\% of the dark matter having {\em arbitrarily strong interactions} if the other 90\% does not interact at all. Even more may be allowed. One reason that this is plausible is that we already know that 15\% of all matter interacts strongly and forms dense structures not captured by simulations of dark matter halos, without violating the constraints from observation: we refer, of course, to ordinary baryonic matter. In fact, it is argued that baryon condensation at the center of the halo could improve the agreement between observations of halo shape and cold DM simulations~\cite{Dubinski:1991bm, Kazantzidis:2004vu, Debattista:2007yz, Valluri:2009ir}; DDDM could have a similar, numerically smaller, effect.

Observations of the Bullet Cluster have also set bounds on dark matter self-interactions~\cite{Markevitch:2003at,Randall:2007ph}, which for SIDM have been argued to be weaker than those from halo shapes~\cite{Feng:2009mn,Peter:2012jh}. They are, however,  more readily interpreted as a bound on PIDM than the halo shape bounds. In the Bullet Cluster, two merging clusters have led to separation between collisional material (hot gas) and collisionless material (stars and ordinary dark matter). The mass of dark matter in the subcluster with stars (inferred from gravitational lensing) leads to the conclusion that no more than 30\% of the dark matter has been lost to collisional effects. Thus we expect that, if the bulk of dark matter is completely collisionless, the Bullet Cluster bound tells us that a subdominant component making up 30\% of all dark matter can have arbitrarily strong self-interactions. It would be left behind with the gas, without changing the lensing observation that tells us the dominant component of dark matter moved with the collisionless stars.

In fact, the most stringent PIDM abundance constraint arises only when there is dissipation and a disk is formed. In that case, stellar velocities in and out of the plane of the galaxy yield stronger bounds. Such velocity distributions offer interesting prospects for reconstructing the galaxy's gravitational potential and hence inferring the distribution of matter within it. The most relevant bounds to date come from the Oort limit, i.e. the inferred local density of matter near the Sun from observations of nearby stars. A recent determination that the local dark matter density is $0.3\pm0.1~{\rm GeV/cm}^3$~\cite{Bovy:2012tw} relied on the kinematics of stars between 1 and 4 kpc above the galactic plane. Another recent determination obtained a similar, but slightly larger, density $0.43\pm0.11\pm0.10~{\rm GeV/cm}^3$~\cite{Salucci:2010qr}. Older results were based on stars within 100 pc of the Sun, surveyed by the Hipparcos satellite, and another sample of stars extending out to 1 kpc~\cite{Holmberg:2004fj,Weber:2009pt}. In the presence of a possible dark disk~\cite{Read:2008fh} these observations were estimated to be consistent with a local dark matter density between $0.2$ and $0.7~{\rm GeV/cm}^3$~\cite{Weber:2009pt}. For our purposes, the most convenient form of the bound is the constraint on the surface density measured below a height $z_0$, which is defined by:
\beq
\Sigma(\left|z\right| < z_0) \equiv \int_{-z_0}^{z_0} \rho(z) dz.
\eeq
$\Sigma$ is approximately equal to the vertical gradient of the gravitational potential, $(2\pi G_N)^{-1} \partial_z \Phi$, which determines the vertical acceleration of stars. As quoted in~\cite{Weber:2009pt}, the total surface density inferred from stellar kinematics is $\Sigma_{\rm tot}(\left|z\right|<1.1~{\rm kpc}) = 71 \pm 6~M_\odot/{\rm pc}^2$. The surface density inferred from visible baryonic matter (stars, stellar remnants, and interstellar gas) is $\Sigma_{\rm vis} = 35~{\rm to}~58~M_\odot/{\rm pc}^2$. We interpret the difference between these numbers as an approximate measure of the amount of DDDM allowed by data. The ranges are one sigma error bars, from which we conclude that at 95\% confidence level the amount of surface density in nonbaryonic matter is
\beq
\Sigma_{\rm dark}(\left|z\right| < 1.1~{\rm kpc}) \simlt 46~M_\odot/{\rm pc}^2.
\label{eq:surfacedensitybound}
\eeq
For the distribution of matter within the disk we use the isothermal sheet model (see e.g. section 11.1 of Ref.~\cite{MvdBW}). If the total mass of DDDM in the galactic disk is $\epsilon M^{\rm gal}_{\rm DM}$, we approximate the volume distribution of DDDM as
\beq
\rho(R,z) = \frac{\epsilon M^{\rm gal}_{\rm DM}}{8\pi R_d^2 z_d} \exp(-R/R_d) {\rm sech}^2(z/2z_d).
\label{eq:distribution}
\eeq
Here $z$ parameterizes height above the midplane of the disk, while $R$ is the radial direction within the disk. We assume the DDDM disk has a scale radius comparable to that for baryons, $R_d \approx 3$ kpc~\cite{BinneyTremaine}. The value of $R$ relevant for the measurements is the distance of the Sun from the galactic center, about 8 kpc. We will discuss the expected values of the disk scale height $z_d$ in Section~\ref{sec:diskform}. For now, we only need to assume $z_d \ll 1.1~{\rm kpc}$, in which case the surface density does not depend on $z_d$:
\beq
\Sigma_{\rm disk}(\left|z\right| < 1.1~{\rm kpc}) = \frac{\epsilon M^{\rm gal}_{\rm DM}}{2\pi R_d^2} \exp(-R/R_d).
\eeq
Given this functional form, from the the surface density bound, Eq.~\ref{eq:surfacedensitybound} is a constraint on the fraction of all the dark matter that is allowed to be in a thin disk:
\beq
\epsilon \simlt 0.05.
\eeq
This is a key result of our paper: the mass of the DDDM disk can be on the order of five percent of the total mass of the Milky Way. Up to order-one uncertainties, this means the mass of the DDDM disk can be as large as the mass of the baryonic disk, and that DDDM can carry comparable energy density to ordinary baryonic matter. It will be very interesting to explore whether improved measurements could detect new structures like DDDM disks. For instance, the ambitious plans of the Gaia satellite (see~\cite{Famaey:2012ga} and references therein) to produce an extensive map of the kinematics of a billion objects in the Milky Way could lead to a powerful probe of dark structures within the galaxy.

Other bounds can in principle arise from bounds on compact objects. Once a sufficiently cold disk has formed, further structure can develop within the disk. Depending on details of DDDM chemistry and molecular cooling that are difficult to calculate, these structures could range from large gaseous clouds down to ``DDDM non-nuclear-burning stars'' that radiate dark photons as the matter within them annihilates. We can estimate the size of large clouds that form within the cold disk based on the Jeans mass, where we treat the clouds as a monatomic gas with sound speed $\sqrt{\frac{5T}{3m}}$:
\beq
M_J & = & \frac{\pi}{6} \left(\frac{5\pi T}{3 G_N m}\right)^{3/2} \left(\frac{1}{\rho}\right)^{1/2} \nonumber \\
& \approx & 10^5 M_\odot \left(\frac{100~{\rm GeV}\times T}{m\times 10^4~{\rm K}}\right)^{3/2} \sqrt{\frac{1~{\rm GeV/cm}^3}{\rho}}.
\eeq
Of course, once clouds above the Jeans mass begin to collapse, atomic and molecular cooling processes could lead to
formation of much smaller structures.

We summarize existing bounds on MACHOs (Massive Compact Halo Objects). For the largest structures above $10^6 M_\odot$, constraints arise from heating of the disk by gravitational scattering of stars on MACHOs~\cite{Lacey:1985}. For structures above about 100~$M_\odot$, including Jeans-scale clouds, the best constraints arise because MACHOs could disrupt wide binary star systems~\cite{Yoo:2003fr}. Smaller objects below 100~$M_\odot$ are constrained by microlensing surveys such as MACHO, EROS, and OGLE. These surveys are reviewed in Ref.~\cite{Moniez:2010zt} and some of their implications for dark matter are discussed in Ref.~\cite{Iocco:2011jz}. Observations have looked toward the Magellanic clouds, which are at relatively high galactic latitude and not likely to constrain DDDM. Other observations toward the galactic center could be more interesting. In the case of ordinary dark matter distributed throughout the halo, for a wide range of masses the MACHO bounds exclude the possibility that more than 10 to 15\% of the halo consists of objects of a given mass~\cite{Yoo:2003fr}. However, we have to be careful when applying these bounds to DDDM, since compact DDDM objects are localized in a disk that can be thinner than the baryonic disk and furthermore is unlikely to be precisely aligned. Therefore the limits on larger objects that derive from the interaction of baryonic matter and dark matter, even if purely gravitational interactions are assumed, need not apply. For objects smaller than the DDDM Jeans mass, we clearly cannot say anything without a more detailed understanding of substructure.

We conclude that while MACHO-type bounds might ultimately detect or constrain our scenario, there is no hard limit at present. The Oort limit is easier to interpret as a bound on DDDM, and we take it as our sole constraint on $\epsilon$. Perhaps future work on compact DDDM objects, together with new analyses of existing data, could lead to stronger bounds or detection prospects.

%%%%%%%%%%%%%%%%
\section{Early Thermal History (Before Structure Formation)}
\label{sec:history}
%%%%%%%%%%%%%%%%%%%%%%%%%%%%%%%%%%%%%%%%

For the purposes of this paper, we  consider a simple PIDM model, with a new abelian gauge group U(1)$_D$, with fine structure constant $\alpha_D$, interacting with two matter fields: a heavy fermion $X$ and a light fermion $C$ (for ``coolant,'' as we will see in Section~\ref{sec:cooling}), with opposite charges $q_X = +1$ and $q_C = -1$ under U(1)$_D$. Of course, one could also consider the case that $X$ and $C$ are scalars. An interesting generalization that we will discuss briefly is the possibility that the new gauge group is nonabelian, SU$(N)_D$, with $X$ in the fundamental and $C$ in the antifundamental representation. In the nonabelian case we make a self-consistent assumption that the confinement scale is far below the temperatures relevant for the phenomena we study. (In the SU(2) case, one can introduce a global symmetry to play an analogous role to the distinction between fundamental and antifundamental representations.)

In this section we will discuss the thermal history of the dark sector, including the amount of dark radiation and the abundance of $X, {\bar X}$ and $C, {\bar C}$ particles. The result is that the predicted dark radiation is allowed by current bounds on the number of relativistic degrees of freedom at the time of BBN and of the CMB, but large enough that Planck can see an interesting signal.

We will also show that the thermal relic abundance of $X$ and ${\bar X}$ can be of the order of the Oort limit, comparable to baryon density.  This  relic symmetric population of $X$ and ${\bar X}$  can annihilate and provide an indirect detection signal.
However, the light particles $C$ and ${\bar C}$ annihilate away efficiently in the early universe. We therefore  have the additional requirement  that there is a nonthermal asymmetric abundance of  $X$ and $C$ that survives to late times, analogous to the nonthermal abundance of protons and electrons in the SM.

\subsection{The Temperature of the Dark Sector}
\label{subsec:temp}

The light degrees of freedom in our scenario introduce constraints (or possible signatures) since at early times they were relativistic and affected the expansion rate of the universe. At the time of BBN, the thermal bath of dark photons and also of the light species $C, {\bar C}$ will add to the total amount of relativistic energy density. At the time of last scattering in the visible sector, only the dark photons will be relativistic. The bounds from BBN and the CMB on relativistic degrees of freedom are usually phrased in terms of the number of effective neutrino species, so we will now calculate the expected number of effective neutrino species present in our model, assuming a sufficiently high decoupling temperature that we justify in Appendix~\ref{app:models}.

Suppose that, at early times, the DDDM sector and the Standard Model were in thermal equilibrium. After decoupling, the entropy density should be separately conserved in the visible and dark sectors. This means that
\beq
\frac{g_{*s,D}^{\rm dec} }{g_{*s,D}(t) \xi(t)^3} =
\frac{g_{*s,{\rm vis}}^{\rm dec}}{g_{*s,{\rm vis}}(t)}
\eeq
with $\xi \equiv \left( T_D/T_{\rm vis}\right)$ and $g_{*s}$ the effective number of degrees of freedom contributing to entropy density. The subscript $D$ refers to dark sector degrees of freedom. Note that $\xi$ is, in general, a time-dependent quantity, as (for example) the visible sector temperature will increase relative to the dark sector temperature whenever visible degrees of freedom decouple from the thermal bath. Suppose that decoupling of the hidden and visible sectors occurs at temperatures below the $W$ mass but above the $b$-quark mass, which is the case if all the mediator particles have weak-scale masses. At this time, $g_{*s,{\rm vis}}^{\rm dec} = 86.25$. The dark plasma, at this time, will contain the dark photons and $C, {\bar C}$ particles, leading to $g_{*s,D}^{\rm dec} = 2 + \frac{7}{8} \times 4 = 5.5$. It is also interesting to consider the generalization to an SU$(N)$ dark sector with $C$ in the fundamental representation, for which $g_{*s,D}^{\rm dec}(N) = 2\left(N^2 - 1\right) + \frac{7}{2}N$. In the visible sector, at the time of BBN we take $g_{*s,D}^{\rm BBN} = 10.75$, while we expect the dark sector degrees of freedom to be unchanged. This leads to
\beq
\xi(t_{\rm BBN}) = \left(\frac{10.75}{86.25}\right)^{1/3} \approx 0.5.
\eeq
The number of additional effective neutrino species is determined by $g_{*s,D} \xi^4(t_{\rm BBN}) = \frac{7}{8} \times 2 \times \Delta N^{\rm BBN}_{{\rm eff},\nu}$, leading to:
\beq
\Delta N^{\rm BBN}_{{\rm eff},\nu} & = & 0.20~{\rm for~U(1)}_D~{\rm and}\nonumber \\
\Delta N^{\rm BBN}_{{\rm eff},\nu} & = & 0.07 N^2 + 0.12 N - 0.07~{\rm for~SU}(N)_D.
\eeq
Numerically, $\Delta N^{\rm BBN}_{{\rm eff},\nu}$ is 0.46 in the SU(2)$_D$ model, 0.94 in the SU(3)$_D$ model, and 1.56 in the SU(4)$_D$ model. Ref.~\cite{Cyburt:2004yc} derives a conservative bound on extra-degrees of freedom during BBN,
\beq
\Delta N_{{\rm eff}, \nu}^{\rm BBN } < 1.44 \ \ {\rm at}\ \ 95\%\ {\rm C.L.},
\eeq
so the U(1)$_D$ model is easily safe. The SU$(N)_D$ model satisfies the bound for $N \leq 4$, with $N=4$ barely outside the 95\% confidence region but easily inside if we assume decoupling at temperatures above the top quark mass when $g^{\rm dec}_{*s,{\rm vis}} = 106.75$. For an alternative point of view, we can relax our assumption about the decoupling temperature and ask: for what value of $g^{\rm dec}_{*s,{\rm vis}}$ is the BBN constraint saturated? It turns out that as long as
\beq
g_{*s,{\rm vis}}^{\rm dec} > 19.3
\eeq
the bound is satisfied for the abelian model. This is the number of degrees of freedom when $T_{\rm vis}^{\rm dec} \approx 200$~MeV.

An equally significant bound on the number of radiation degrees of freedom comes from the CMB. A recent analysis of nine years of WMAP data~\cite{Hinshaw:2012fq} combined with the terrestrial experiments SPT~\cite{Keisler:2011aw} and ACT~\cite{Das:2011ak} and baryonic acoustic oscillations constrains $\Delta N_\nu^{\rm CMB} < 1.6$ at 95\% C.L. Very recently, the Planck Collaboration has published a stronger bound~\cite{Ade:2013lta}:
\beq
\Delta N_\nu^{\rm CMB} < 1.0 \ \ {\rm at} \ \ 95\%\ {\rm C.L.},
\eeq
using the ``Planck+WP+highL+$H_0$+BAO'' result in which the Hubble scale floats in the fit. At the time of last scattering in the visible sector, we have $g^{\rm CMB}_{*s,{\rm vis}} = 3.36$ (from photons and the colder neutrinos) and $g_{*s,D}^{\rm CMB} = 2$ (from dark photons) or $2(N^2-1)$ (in the nonabelian case). At this time the temperature ratio is
\beq
\xi &=& \left(\frac{5.5}{2} \times \frac{3.36}{86.25}\right)^{1/3} \approx 0.5~{\rm for~U(1)}_D,\nonumber \\
\xi &=& \left(\frac{2\left(N^2 - 1\right) + \frac{7}{2}N}{2\left(N^2 - 1\right)} \times \frac{3.36}{86.25}\right)^{1/3}~{\rm for~SU}(N)_D,
\eeq
Robustly, if the two sectors are in thermal equilibrium near the weak scale, we expect the dark photon temperature to be around half the visible photon temperature. Alternative cosmologies, for instance with decoupling at much higher temperatures below which many new visible-sector degrees of freedom exist, could allow much smaller $\xi$, but we will generally take $\xi \approx 0.5$ throughout the paper.

The temperature of dark recombination (formation of dark atoms from $X$ and $C$ ions) is about a factor of ten below the binding energy $B_{XC} \sim \alpha_D^2 m_C$. Large $\alpha_D$ suppresses the thermal relic abundance of $X, {\bar X}$ and larger $m_C$ prevents efficient cooling, as we will see in Section~\ref{sec:cooling}. Hence, we favor parameter space at small $B_{XC}$ where recombination in the dark sector doesn't happen until after last scattering in the visible sector.  This means that when the CMB is formed, dark photons are interacting with the dark fluid of $X$ and $C$ particles, with a speed of sound slightly less than, but of order, $c/\sqrt{3}$. Much as for ordinary baryons, there will be dark acoustic oscillations and other effects from this nontrivial coupling of radiation to matter. Although not entirely correct since the additional degrees of freedom in our model are not yet free streaming, we interpret the bound on the number of effective neutrino species as a bound on free dark photons, ignoring the coupling to the fluid. We expect that, because the sound speed is of order the speed of light, this will be a good approximate guideline to whether the theory is allowed by the current data. It will be very interesting to do a more careful analysis that can distinguish this scenario.

 The number of additional effective neutrino species is determined by $g_{*s,D} \xi^4(t_{\rm CMB}) = \left(\frac{4}{11}\right)^{4/3} \times \frac{7}{8} \times 2 \times \Delta N^{\rm CMB}_{{\rm eff},\nu}$, leading to:
\beq
\Delta N^{\rm CMB}_{{\rm eff},\nu} &=& 0.22~{\rm for~U(1)}_D, \nonumber \\
\Delta N^{\rm CMB}_{{\rm eff},\nu} &=& 4.4(N^2-1)\xi^4~{\rm for~SU}(N)_D.
\eeq
Numerically, $\Delta N^{\rm CMB}_{{\rm eff},\nu}$ is 0.49 in the SU(2)$_D$ model, 0.91 in the SU(3)$_D$ model, and 1.45 in the SU(4)$_D$ model, so the bound is satisfied for $N < 4$. In the abelian model, if we ask what value of $g^{\rm dec}_{*s,{\rm vis}}$ saturates the CMB bound, we find it is
\beq
g_{*s,{\rm vis}}^{\rm dec} > 28.1,
\eeq
a slightly tighter bound than we derived from BBN. Thus, the abelian model is allowed provided the two sectors decoupled at temperatures above the QCD phase transition. The SU$(N)_D$ is allowed for $N \leq 4$, and predicts sizable deviations in the number of effective neutrino species. Further analysis of Planck data in combination with other experiments may help to clarify the number of relativistic species at the time of the CMB~\cite{Perotto:2006rj,Hamann:2007sb,Joudaki:2012fx}. Improved measurements of $\Delta N^{\rm CMB}_{{\rm eff},\nu}$ will also come from ACTpol~\cite{Niemack:2010wz} and SPTpol~\cite{Austermann:2012ga}. Finally, we note that related comments on the number of allowed dark gauge bosons appeared recently in Ref.~\cite{Franca:2013zxa}.

\subsection{Relic Abundance of $X$ and $C$}
\label{subsec:relicabundance}

\begin{figure}[!h]\begin{center}
\includegraphics[width=0.45\textwidth]{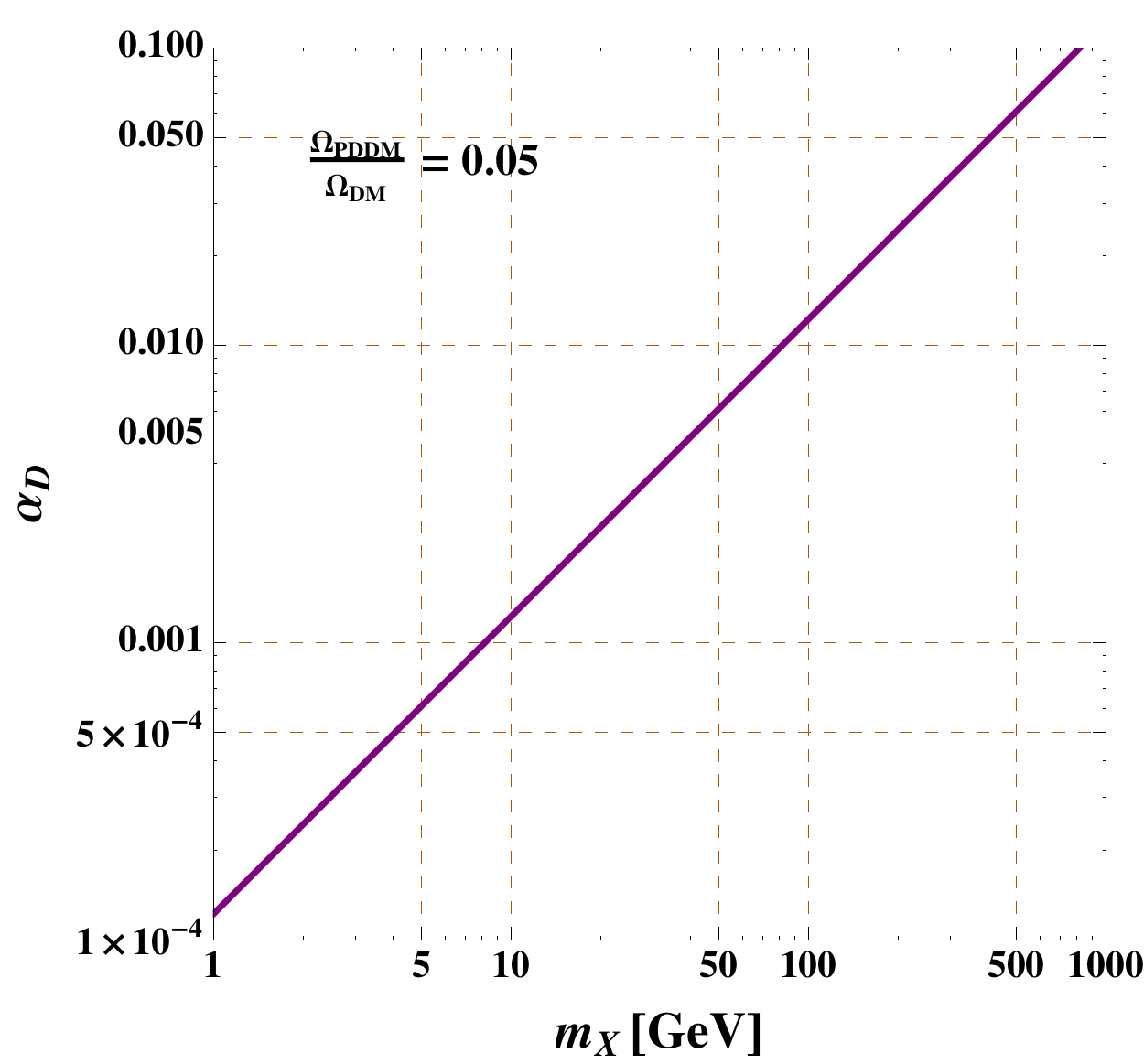}
\end{center}
\caption{$\alpha_D$ that yields a thermal relic abundance of $X, {\bar X}$ that is a 5\% fraction of the total DM density for different $m_X$.}
\label{fig:relic}
\end{figure}

Having considered the relic radiation, we now consider the relic abundances of $X$ and $C$. The thermal relic abundance of particles charged under a hidden U(1)$_D$ has been discussed in Refs.~\cite{Feng:2008mu,Ackerman:2008gi,Feng:2009mn}. Depending on whether the mediator particles coupling $X$ to the Standard Model thermal bath are heavier or lighter than $X$, the dark sector may be at precisely the same temperature as the SM when $X$ freezes out, or as we saw in the previous subsection it could have about half the Standard Model temperature if the two sectors have decoupled. In Figure~\ref{fig:relic}, we have plotted the curve in the $(m_X, \alpha_D)$ plane which predicts $\epsilon = 0.05$ for the thermal relic abundance of $X$ and ${\bar X}$, assuming the SM and the hidden sector are still at the same temperature at the time of $X$ decoupling. The relic abundance was calculated via the standard analytic formula (Eq. (33) - (34) in Ref~\cite{Feng:2008mu}). Taking the hidden sector to be at half the SM temperature leads to slightly lower values of $\alpha_D$. Values of $\alpha_D$ below the line in Fig.~\ref{fig:relic} lead to a large relic abundance that violates the Oort limit discussed in Section~\ref{sec:pddmamount}, whereas larger values of $\alpha_D$ are allowed  only with a nonthermal mechanism for generating more $X$ particles.

%%%%%%%%%%%%%%%%%%%%%%%%%%%%%%%%%%%%%%%%%%%%%%%%%%%%%%%
%%%%%%%%%%%%%%%%%%%%%%%%%%%%%%%%%%%%%%%%%%%%%%%%%%%%%%

\begin{figure}[t]
\centering
\includegraphics[width=0.42\textwidth]{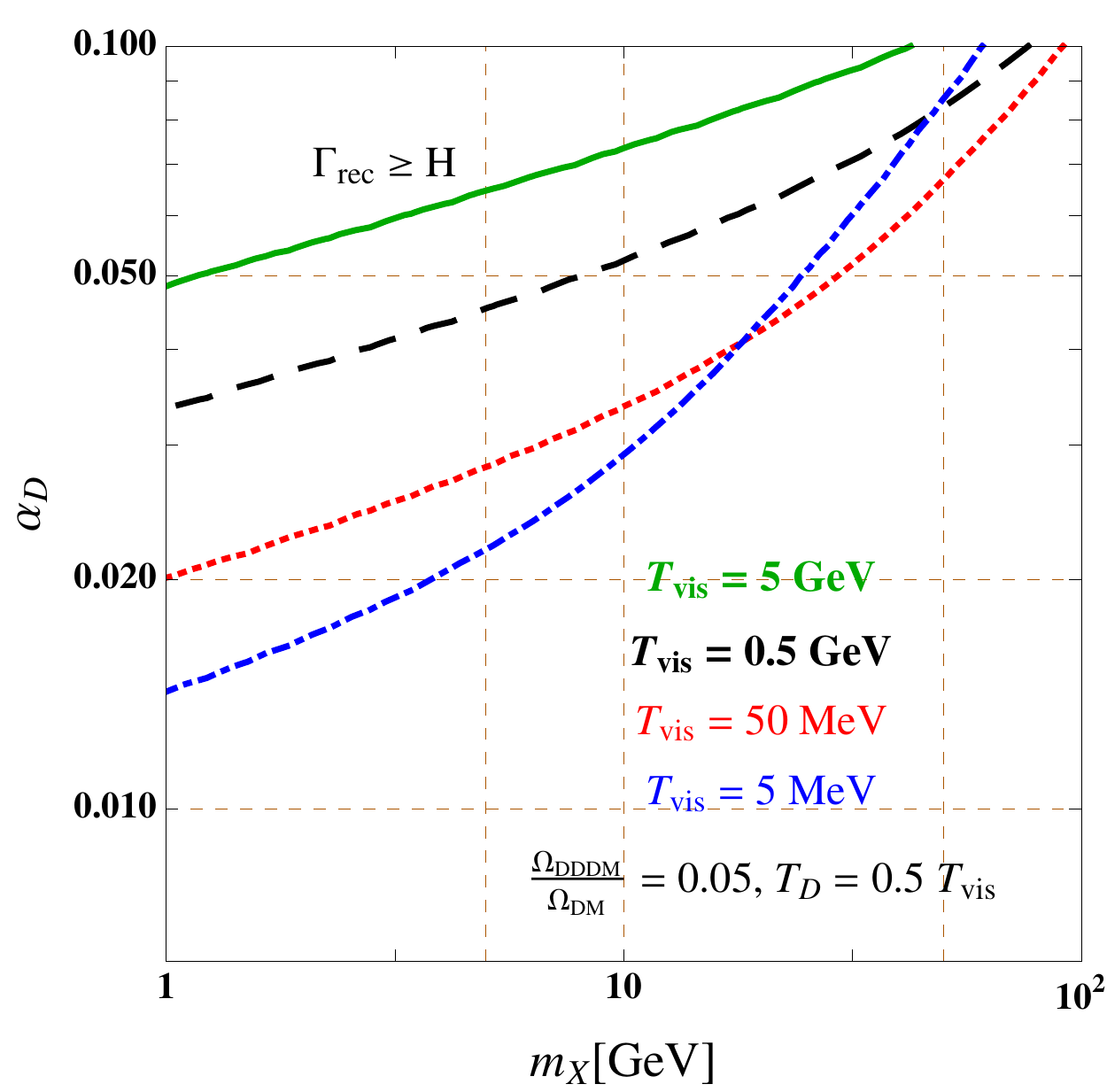}
\caption{Above the curves, the recombination rates are larger than the Hubble rate, leading to $X{\bar X}$ annihilation that depletes the abundance of the symmetric relic component.}
\label{fig:recomb}
\end{figure}

Figure~\ref{fig:relic} clearly shows that we can achieve a relic density of $X$ and ${\bar X}$ particles that saturates the Oort bound for reasonable values of the coupling $\alpha_D$. It is possible that the relic density estimated at freezeout is later decreased by two processes with similar rates: Sommerfeld-enhanced annihilation of $X$ and ${\bar X}$ at low temperatures and recombination into $X{\bar X}$ bound states which annihilate away. Using the recombination rate given in~\cite{Bethe:1934za,Feng:2009mn}, we plot in Fig.~\ref{fig:recomb} the curves $\Gamma_{\rm rec} = H$ for particular choices of visible and dark sector temperatures. These show that recombination and subsequent annihilation of $X{\bar X}$ bound states does not wash out the abundance of $X, {\bar X}$ for $\alpha_D \simlt 0.01$ and $m_X \simgt 1$ GeV.

%%%%%%%%%%%%%%%%%%%
%%%%%%%%%%%%%%%%%%%

The light species $C$ with $m_C \ll m_X$ freezes out at much later times, and has a much larger annihilation rate than the heavy species, by a factor $\left(m_X/m_C\right)^2$. As a result, the thermal relic number density of $C$ is much smaller than that of $X$, by a factor $m_C/m_X$. This means that we expect any symmetric component of $C$ and ${\bar C}$ to annihilate away almost completely at dark sector temperatures a factor of 20 below the $C$ mass. The existence of light $C$ particles is crucial to dissipative dynamics, as we will see in detail in Section~\ref{sec:cooling}. This means that only a nonthermal mechanism for producing $C$ particles can be consistent with dissipative dynamics.

 We assume asymmetric nonthermal abundances of DDDM, in which we have a net $C$ number $n_C - n_{\bar C} \neq 0$. The universe should be charge symmetric and this means that there should also exist an asymmetry in $X$, $n_X - n_{\bar X} = n_C - n_{\bar C}$. We  assume that $n_{\bar X} = n_{\bar C} = 0$ at late times, whereas $X$ and $C$ survive. The idea of asymmetric dark matter has inspired many proposals for how the dark matter and baryon abundances may be related~\cite{Nussinov:1985xr,Kaplan:1991ah,Kaplan:2009ag,Kribs:2009fy,Shelton:2010ta,Davoudiasl:2010am,Buckley:2010ui,Falkowski:2011xh,MarchRussell:2011fi,Cui:2011qe,Cui:2011ab}, and similar considerations could explain the amount of DDDM. We expect for reasonable parameters that the asymmetric abundance of $X$ and $C$ will be supplemented with a relic symmetric population of $X$ and ${\bar X}$, allowing for the prospect of interesting indirect detection signals when $X$ and ${\bar X}$ annihilate to Standard Model particles.

It is possible that other nonthermal scenarios could leave a symmetric population of $C$ and ${\bar C}$ that survives to late times. For instance, late-decaying dark sector particles could produce additional $C$ particles after thermal freezeout of $C$. However, even these late-produced $C$ particles could annihilate at temperatures below $B_{C{\bar C}} = \alpha_D^2 m_C/4$, through Sommerfeld-enhanced annihilation or recombination. One way to prevent this would be if they were first bound up into dark atoms with $X$, which have a larger binding energy and might protect the $C$ particles from annihilation. However, recombination of $XC$ dark atoms is much slower than naively expected, because one dark atom can be ionized by the dark photon emitted when another dark atom is formed. This is analogous to what happened for ordinary hydrogen in our universe~\cite{Peebles:1968ja}. It has been studied for dark atoms in Refs.~\cite{Kaplan:2009de,CyrRacine:2012fz}. The result is that $C{\bar C}$ annihilation would always be much faster than the sequestering of $C$ particles inside dark atoms with $X$.

A loophole arises in the case of nonabelian gauge theories. In that case, the dark gluon emitted by formation of one dark atom could ionize another dark atom, but it could also first encounter another dark gluon and scatter off it. In the process, one of the two dark gluons loses energy. Because the number of formed atoms and the number of dark gluons emitted in recombination must be equal, we expect an order-one fraction of dark gluons can lose energy and thus become too soft to ionize a dark atom. This could allow the $C$ and ${\bar C}$ particles to be stored in tightly-bound $XC$ and ${\bar X}{\bar C}$ bound states, possibly allowing an interesting symmetric abundance of $C,{\bar C}$ to survive to late times and contribute to cooling. A full numerical analysis of this rather complicated cosmology is beyond the scope of this paper.

 Because the cosmology in which we simply assume an asymmetric component of $X$ and $C$ is  simpler,  through most of the paper we will assume asymmetric dark matter.

%%%%%%%%%%%%%%%%%%%%%%%%%%%%%%%%%%%%%%%%
 \section{The Fermi Line and Other Indirect Detection Signals}
 \label{sec:fermiline}
 %%%%%%%%%%%%%%%%%%%%%%%%%%%%%%%%%%%%%%%%

 The DDDM scenario was motivated in part by the observation of gamma-ray line emission at about 130 GeV (135 GeV after energy recalibration) in the galactic center using 3.7 years of Fermi-LAT data~\cite{Bringmann:2012vr,Weniger:2012tx,Tempel:2012ey,Su:2012ft}. The signal has also been claimed to exist in galaxy clusters~\cite{Hektor:2012kc}. The observation is not decisive, and certain features in the data suggest it could very well be an instrumental effect~\cite{Hektor:2012ev,Finkbeiner:2012ez,Whiteson:2012hr,Whiteson:2013cs}. Nonetheless, the  suggestion serves as a concrete example where a possible enhancement of an indirect signal inherent in our scenario could be critical. In fact, {\em any} high-energy gamma ray line observable in the near future and consistent with continuum bounds would require similar phenomenology. Furthermore, a large flux of high-energy electrons and positrons recently observed in cosmic rays by PAMELA and subsequent experiments~\cite{Adriani:2008zr,Chang:2008aa,Abdo:2009zk,Aharonian:2009ah,FermiLAT:2011ab} also requires large boost factors to explain in terms of dark matter annihilation~\cite{Cholis:2008hb} (though it is plausibly due to pulsars). This shows that, even apart from the specific case of the Fermi-LAT line, it is very interesting to consider general mechanisms that can produce large boost factors. We will see in this paper that boost factors as large as 10,000 are conceivable for DM of mass 130 GeV with U(1)$_D$ coupling in the range to give the right thermal relic density. The large astrophysical enhancement we find will be due to a very thin dark matter disk. In modified models, related enhancements could explain boost factors needed for PAMELA or a possible AMS signal~\cite{Palmonari:2011zz} as well.

 We first consider the required signal enhancement in the context of a simple model in which DM, $\phi$ and the new charged particles $S$ are all scalars.
 \beq
-{\cal L} & \supset & \lambda_{\phi S} |\phi|^2 |S|^2 + m_{S}^2 |S|^2 + \lambda_S |S|^4 + m_\phi^2 |\phi|^2 + \lambda_\phi |\phi|^4, \nonumber
\eeq
 in which $\phi$ is charged only under U(1)$_D$ and $S$ has charge 1 under the usual U(1)$_{\rm EM}$. One could also consider models with fermionic DM. In that case, to avoid kinetic mixing between U(1)$_D$ and U(1)$_{\rm EM}$~\cite{Holdom:1985ag}, one could design a resonant annihilation model where DM annihilates through an intermediate boson and a U(1)$_{\rm EM}$ charged particle loop to two photons, although anomaly-like constraints on charge assignments can make such models safe (as explained in Appendix~\ref{app:models}) even when a particle is charged under both U(1)s.

\begin{figure}[!h]\begin{center}
\includegraphics[width=0.4\textwidth]{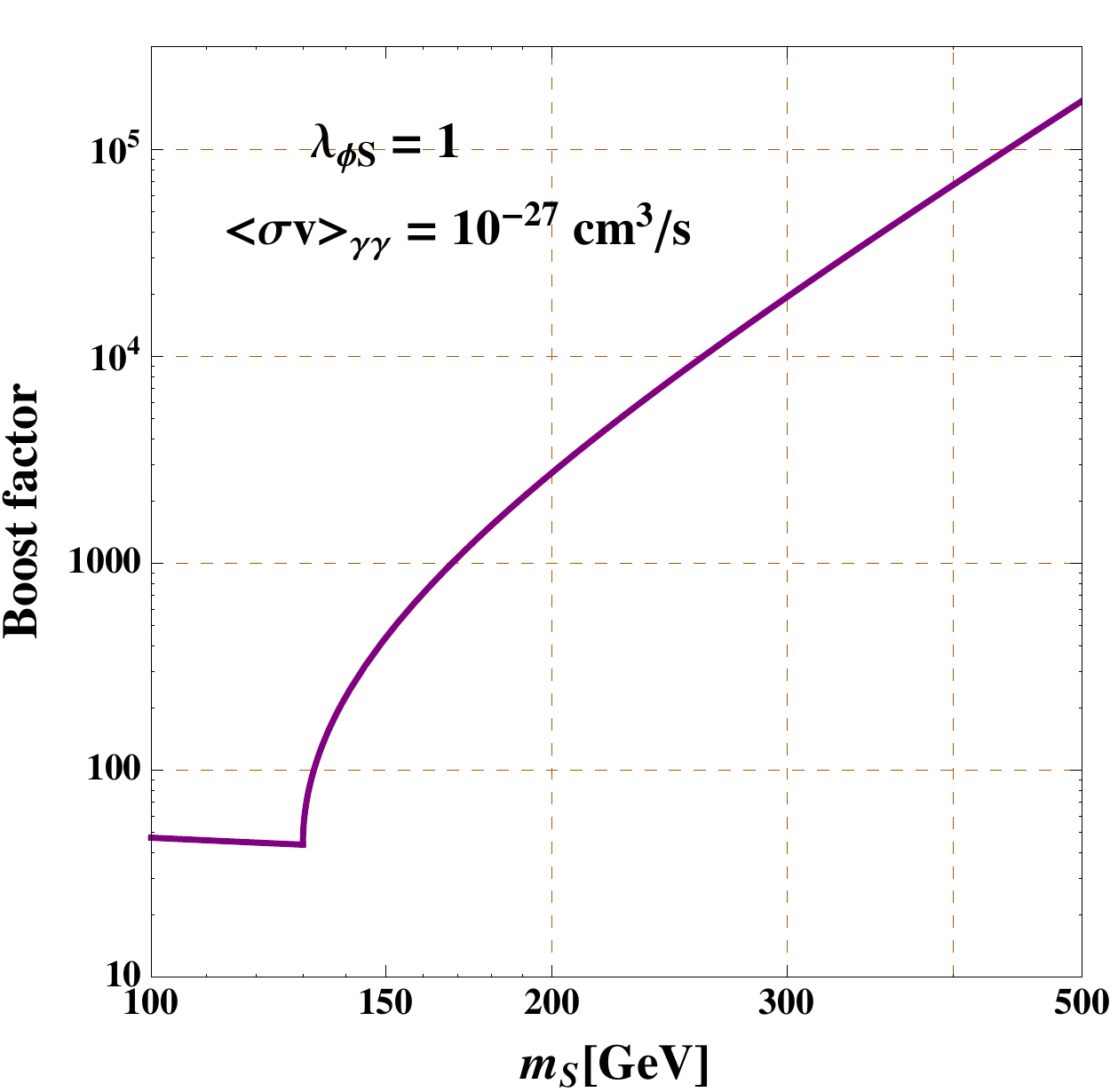}
\end{center}
\caption{Fixing $(\sigma v)( \phi \phi \to \gamma \gamma) = 10^{-27}$ cm$^{3}$s$^{-1}$, $\lambda_{\phi S}=1$ and $m_\phi = 130$ GeV, the boost factor $B$ needed to generate a signal that current Fermi line search is sensitive to.}
\label{fig:scalar}
\end{figure}

The observed photon line could be consistent with DM particles annihilating at one loop to $\gamma\gamma$ with an unexpectedly large cross section of order $\langle \sigma v\rangle \sim 10^{-27}$ cm$^3$/s. In our example model, the cross section of the DM annihilation to diphotons would be
\beq
\sigma v_{\phi^\dagger \phi \to \gamma \gamma}  = \frac{B}{32 \pi^3 m_\phi^2} \left|\alpha \lambda_{\phi S} \tau_\phi^{-1} A_0(\tau_\phi)\right|^2,
\eeq
where $B$ is the boost factor which could either come from microscopic physics such as Sommerfeld enhancement, which we will discuss in this section, or from astrophysics, such as the density enhancement we will discuss in the following sections. Here
\beq
A_0(\tau)&= &- \tau + \tau^2 f(\tau^{-1}) \quad {\rm with}\; \tau_\phi= m_S^2/m_\phi^2, \\
f(x)&=&{\rm arcsin}^2\sqrt{x}.
\eeq
(For $m_S < m_\phi$, it is necessary to analytically continue $f(x)$.) Demanding $(\sigma v)( \phi^\dagger \phi \to \gamma \gamma) = 10^{-27}$~cm$^{3}$s$^{-1}$ and fixing $\lambda_{\phi S}=1$, one can derive the required $B$ for a given $m_S$. The result is presented in Fig.~\ref{fig:scalar}. Thus to explain the line, without relying on large couplings and tuning $m_S$ to be close to the DM mass, one needs a huge boost factor. It is also difficult to make this scenario consistent with a thermal relic abundance. The observed $\sigma v$ to photons is too small, leading to an overabundance of $\phi$ if this is the only annihilation channel. But adding larger annihilation channels, such as $\phi^\dagger \phi \to W^+ W^-$, is typically in tension with the absence of an observed gamma ray continuum. One possible explanation is if $S$ is slightly heavier than $\phi$, so that $\phi \phi^\dagger \to S S^\dagger$ can be an important annihilation mode in the thermal environment of the early universe, but is impossible today~\cite{Griest:1990kh,Tulin:2012uq}.

%%%%%%%%%%%%%%%%%%%%%%%%%%%%%%%%%%%%%%%%%%%%%%%%%%%%%%%
%%%%%%%%%%%%%%%%%%%%%%%%%%%%%%%%%%%%%%%%%%%%%%%%%%%%%%

For DM charged under an unbroken U(1)$_D$, there is indeed an enhancement, the Sommerfeld enhancement,
\beq
S=\frac{2\pi \alpha_D Q_D^2/v}{1-e^{-2\pi \alpha_D Q_D^2/v}},
\eeq
where $v$ is the DM velocity. However for $\alpha_DQ_D^2 \le 0.1$ and $v \approx 10^{-3}$, $S \le 1000$. If one fixes the DM thermal relic $\Omega h^2$ to be 0.11, $\alpha_DQ_D^2 \approx 3 \times10^{-3}$, $S \approx 20$. Thus Sommerfeld enhancement itself is not sufficient enough to get the desired annihilation cross section.

 In what follows, we will see that enhanced density from DDDM  could be sufficient to generate such a large boost factor.
Of course the precise value for the boost will depend on the precise parameters of the dark matter candidate (and some as-yet-unknown astrophysics) as we  discuss below.

%%%%%%%%%%%%%%%%%%%%%%%%%%%%%%%%%%%%%%%%
 \section{Cooling}
 \label{sec:cooling}
 %%%%%%%%%%%%%%%%%%%%%%%%%%%%%%%%%%%%%%%%

The enhanced signals we discuss arise as the result of the interacting component of dark matter collapsing into a disk. We now consider when and how this can occur. This is equivalent to the question of when cooling is sufficiently quick to allow for collapse, so we now investigate the question of how interacting dark matter can cool.

The cooling has many features in common with ordinary baryonic matter. DDDM first adiabatically cools through the expansion of the universe. As with baryons during the formation of a galaxy, the interacting dark matter will already be present (in the primordial overdense region that seeds an early galaxy halo or in progenitor halos that merge into a larger galaxy) and will also accrete onto the galaxy from the intergalactic medium. After virialization through shock heating, baryons cool from different processes: atomic and molecular interactions, Compton cooling, and bremsstrahlung radiation. All of these require light electrons in order to have a sufficiently rapid rate.

The same mechanisms will be required for DDDM: cooling occurs sufficiently rapidly only when a light particle is present that also interacts under the dark U(1) (or more generally, whatever force is relevant). Therefore at the time of the initial accretion, part of the DDDM might be bound into atomic-like states of heavy and light dark matter. As discussed in the previous section, for instance, the simplest model is an asymmetric population of $X$ and $C$, which like ordinary hydrogen can form bound states in the early universe with some residual ionization (as calculated carefully in~\cite{CyrRacine:2012fz}). A relic population of $X$ and ${\bar X}$ may also survive, so the initial conditions will involve a mix of dark atoms and dark ions. However, as we will now see that shock heating will destroy any initially bound atoms, we can consider cooling in this section without determining the exact fraction of bound states in the very early universe.

As dark atoms fall into the overdense region, their gravitational potential energy converts to kinetic energy. Initially they are quite cold, but when falling into the galactic center, particles slow down as they encounter other infalling particles, forming a shock wave which expands outward, containing pressure-supported gas inside~\cite{Binney1977,ReesOstriker1977}. This shock-heating process converts the kinetic energy of the DDDM gas to thermal energy at the virial temperature,
\beq
T_{\rm vir} = \frac{G_N M \mu}{5 R_{\rm vir}} \approx 8.6~{\rm keV} \frac{M}{M^{\rm gal}_{\rm DM}} \frac{\mu}{100~{\rm GeV}} \frac{110~{\rm kpc}}{R_{\rm vir}}.
\eeq
where $M$ stands for the mass of the virial cluster and $\mu = \rho/n$ is the average mass of a particle in the DDDM gas. We have taken a fiducial value for the mass of dark matter in the Milky Way galaxy, $M_{\rm DM}^{\rm gal} = 10^{12} M_\odot$. This is reasonable since the initial density perturbation induces gravitational collapse in the dominant dark matter component for which neither baryons nor the subdominant interacting dark matter should be very relevant. Note that for a virial cluster of the same mass and radius, DDDM will be much hotter than baryonic matter, with a temperature enhanced by $\sim m_X/m_p$. The binding
energy of the ground state of the dark atom is
\beq
B_{XC} \equiv \frac{\alpha_D^2 m_C}{2},
\eeq
less than or of order the binding energy of ordinary hydrogen, so we expect $T_{\rm vir } \gg B_{XC}$. At these temperatures the DDDM
in the virial cluster will be completely ionized, even if it had recombined
into dark atoms or dark molecules before virialization. Hence we can start off thinking of free $X$ and $C$ particles.

\begin{figure}[!h]\begin{center}
\includegraphics[width=1.0\textwidth]{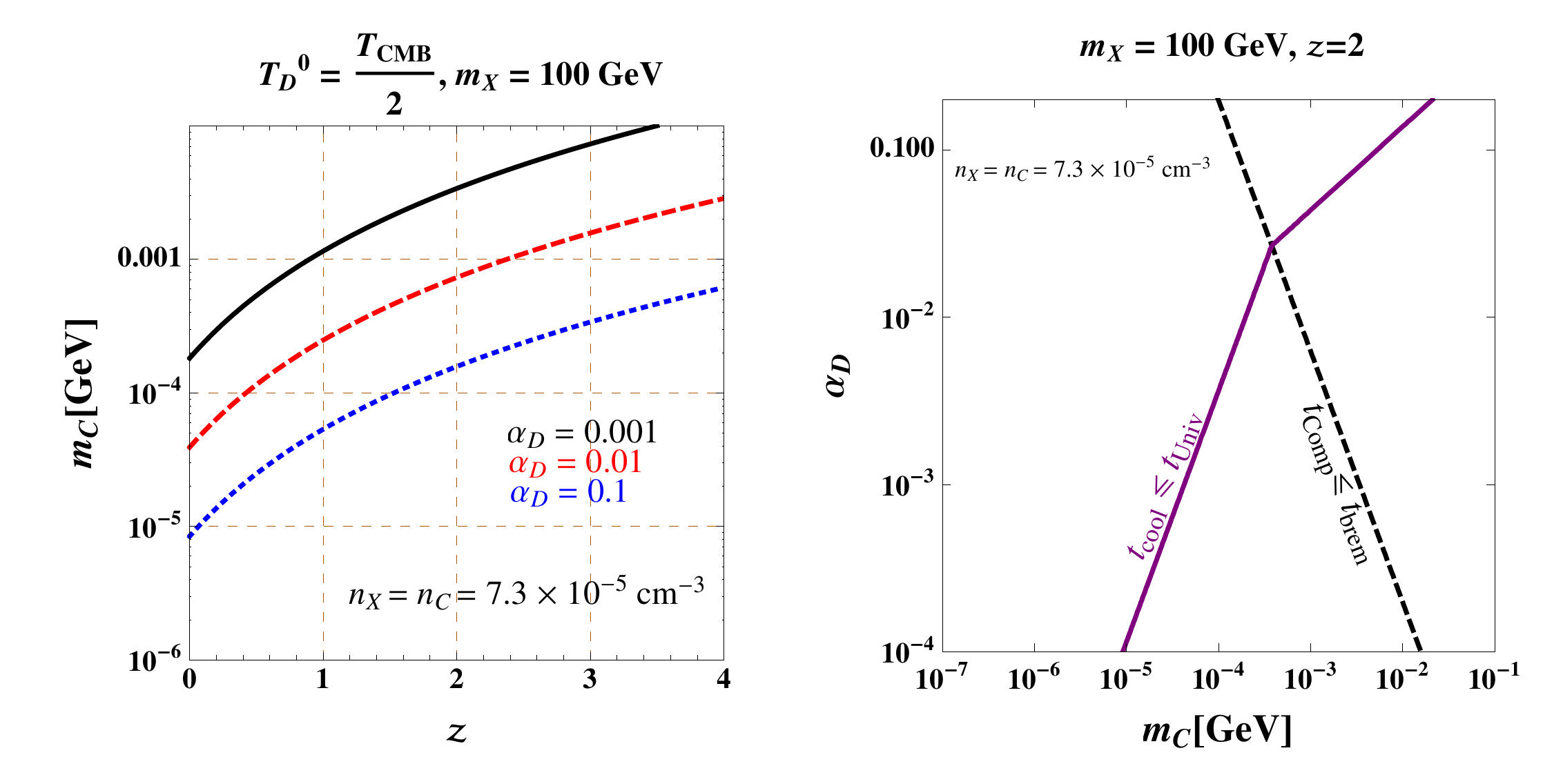}
\end{center}
\caption{Comparison of the rates of bremsstrahlung and Compton cooling. At left: the value of $m_C$ for which the rates are equal, as a function of redshift. To the right of the curves, i.e. at early times, Compton cooling dominates. At right: the contour in the $(m_C, \alpha_D)$ plane along which the bremsstrahlung cooling rate equals the Compton cooling rate (black dashed line) and the contour along which the cooling rate equals the age of the universe (solid purple line). This shows that Compton cooling is the dominant effect at small $m_C$ and $\alpha_D$, while bremsstrahlung dominates for larger values. In both plots, we have taken an NFW virial cluster of radius 20 kpc. }
\label{fig:compton}
\end{figure}

The same cooling processes that apply to baryons potentially apply to DDDM. An ionized dark plasma in the virial cluster can be  cooled through bremsstrahlung and Compton scattering off background dark photons. Compton scattering is more efficient at larger redshift, when the dark photon background was hotter. Based on the results of Section~\ref{subsec:temp}, we take the dark photon temperature to be $T_D \approx 0.5 T_{\rm CMB}$. (This is the temperature of the dark cosmic background photons, which is to be distinguished from the temperature of $X$ and $C$ particles in the galaxy, $T_{\rm vir}$.) The timescale of the bremsstrahlung cooling is
\beq
t_{\rm brem} & \approx & \frac{3}{16} \frac{n_X + n_C}{n_X n_C}
\frac{m_C^{3/2}T_{\rm vir}^{1/2}}{\alpha_D^3}  \\
& \approx & 10^{4}\,{\rm yr}~\sqrt{\frac{T_{\rm vir}}{\rm K}}\frac{{\rm cm}^{-3}}{n_C} \left(\frac{\alpha_{\rm EM}}{\alpha_D}\right)^3 \left(\frac{m_C}{m_e}\right)^\frac{3}{2},\nonumber
\label{eq:tbrem}
\eeq
where in the second line, we assume $n_X = n_C$ for simplicity. At the end of the section, we will relax this assumption. This time should be compared to the age of the universe in order to show that the cluster efficiently cools down. The timescale for cooling through Compton scattering is
\beq
t_{\rm Compton} &\approx & \frac{135}{64\pi^3}  \frac{n_X + n_C}{n_C} \frac{m_C^3}{\alpha_D^2 \left(T_D^0(1+z)\right)^4} \\
&\approx & 4 \times10^{12}\, {\rm yr}~ \frac{n_X + n_C}{n_C} \left(\frac{\alpha_{\rm EM}}{\alpha_D}\right)^2 \left(\frac{2 \,{\rm K}}{T_D^0(1+z)}\right)^4 \left(\frac{m_C}{m_e}\right)^3,\nonumber
\eeq
where $T_D^0$ is the current dark CMB temperature and $z$ is the redshift. In Figure~\ref{fig:compton} at left, we show contours in the plane of $m_C$ and redshift along which the bremsstrahlung and Compton cooling rates are equal, for different choices of $\alpha_D$. Because the Milky Way galaxy was starting to form before $z = 2$, Compton cooling of DDDM would be important within the Milky Way at early times. Compton scattering could also be important for smaller $\alpha_D$ and $m_C$. We illustrate this in the right-hand plot of Fig.~\ref{fig:compton}, which shows the contours in the $(m_C, \alpha_D)$ plane along which the two rates are equal and along which the faster rate equals the age of the universe. As the dark photon background cooled, bremsstrahlung would have become increasingly important. We use the generic term $t_{\rm cool}$ for whichever time scale is shorter: $t_{\rm cool} = \min(t_{\rm brem},t_{\rm Compton})$.

%%%%%%%%%%%%%%%%%%%%%%%%%%%%%%%%%%%%%%%%%%%%%%%%%%%%%%%
%%%%%%%%%%%%%%%%%%%%%%%%%%%%%%%%%%%%%%%%%%%%%%%%%%%%%%

In order to verify that bremsstrahlung or Compton scattering leads to cooling, we first make some consistency checks. The emitted dark photons must escape from the galaxy and carry away energy without being reabsorbed. The primary process by which a dark photon would interact is through scattering with a light $C$ particle, so we can approximate the photon's mean free path by
\beq
\ell = \frac{1}{\sigma_T n_C} = \frac{3 m_C^2}{8\pi \alpha_D^2 n_C} \approx 1.5 \times 10^8~{\rm kpc},
\eeq
where we have used the Thomson cross section for $\gamma_D$--$C$ scattering with $\alpha_D = \alpha$, $m_C = m_e$, and $m_X = 100$ GeV while assuming equal $X$ and $C$ number densities at $\epsilon = 0.05$ and a virial radius of 110 kpc, namely
\beq
n_X = n_C \approx 3.3 \times 10^{-6}~{\rm cm}^{-3} \left(\frac{100~{\rm GeV}}{m_X}\right).
\label{eq:numberdensity110kpc}
\eeq
The long mean free path shows that photons readily escape the galaxy at early times. Furthermore, because $\ell \sim 10^6 R_{\rm vir}$, photons will continue to escape even if the initial DDDM distribution collapses by a factor of $10^{18}$ in volume. This is sufficient to allow a disk to form, especially considering that once the DDDM assumes a disk-like shape, photons can escape more efficiently through the thin direction of the disk.

 We also need to check that both light and heavy particles could cool. When light particles scatter on heavy particles and emit bremsstrahlung photons, it is mostly the light particles that lose energy. Similarly, Compton scattering is dominantly scattering of the light particles on dark background photons. However, if heavy and light particles remain thermally coupled, the cooling of the light particles is sufficient.

  Thermal coupling occurs when the rate for Rutherford scattering of the light particles on the heavy particles exceeds the cooling rate. In this case, the heavy particles cool adiabatically, with scattering keeping the light and heavy species in in kinetic equilibrium~\cite{SpitzerJr.:1941,SpitzerJr.:1942zz}. The timescale for this equilibration process is
\beq
t_{\rm eq}&=& \frac{m_Xm_C}{2\sqrt{3\pi}\alpha_D^2}\frac{(E_C/m_C)^{3/2}}{n_C \log\left(1+\frac{v_C^4m_C^2}{\alpha_D^2n_C^{2/3}}\right)} \\
& = & 4.3 \times 10^4~{\rm yr}  \left(\frac{\alpha}{\alpha_D}\right)^2 \left(\frac{m_X}{1~{\rm GeV}}\right)^\frac{5}{2} \left(\frac{m_e}{m_C}\right)^\frac{1}{2} \frac{{\rm cm}^{-3}}{n_C} \frac{10}{\log\left(1+\frac{v_C^4m_C^2}{\alpha_D^2n_C^{2/3}}\right)},\nonumber
\eeq
where $E_C$ is the kinetic energy of the light species; in the second line, we take $E_C/m_C=3T_{\rm vir}/m_C$. In part of our parameter space, $t_{\rm eq} \ll t_{\rm cool}$ and the light and heavy species  cool adiabatically together.

Rutherford scattering has a $1/v_C^4$ enhancement, but when $m_X/m_C$ is very large, in a thermal system $v_C$ is not small.  Thus for large $m_X/m_C$, as well as in the region of parameter space where $\alpha_D$ is very small, the equipartition time from two-body scattering processes is not sufficient to cool the heavy particles. In this case, we expect cooling should still occur but that cooling involves nonequilibrium physics, at least initially. If the light particles contract as they cool, while the heavy particles are unaffected, a charge separation would occur between the larger cloud of $X$ particles and a smaller cloud of $C$ particles. This wwould produce dark electric fields that pull the $X$ particles in. It would be interesting to simulate or model more completely the resulting dynamics, but it seems inevitable that, since cooling continues to rob the system of kinetic energy, eventually both $X$ and $C$ will cool. As they contract into smaller volume, larger values of $n_{X,C}$ make Rutherford scattering more efficient, and the cooling process will eventually be describable again by equilibrium physics.

Hence, we work under the hypothesis that whenever the cooling time scale $t_{\rm cool}$ is less than the age of the universe, cooling occurs. At this point we should mention one further subtlety: equipartition will speed up the light particles relative to the heavy ones by a factor $\sqrt{m_X/m_C}$, and so for sufficiently small $m_C$ we should use the formula for relativistic bremsstrahlung rather than Eq.~\ref{eq:tbrem}. Since the rate of energy loss from relativistic bremsstrahlung exceeds that from nonrelativistic bremsstrahlung by a Lorentz factor, cooling will  become only faster~\cite{Bethe:1934za}. Thus, Eq.~\ref{eq:tbrem} is a conservative estimate.

%%%%%%%%%%%%%%%%%%%%%%%%%%%%%%%%%%%%%%%%%%%%%%%%%%%%%%%
\begin{figure}[!h]\begin{center}
\includegraphics[width=0.97\textwidth]{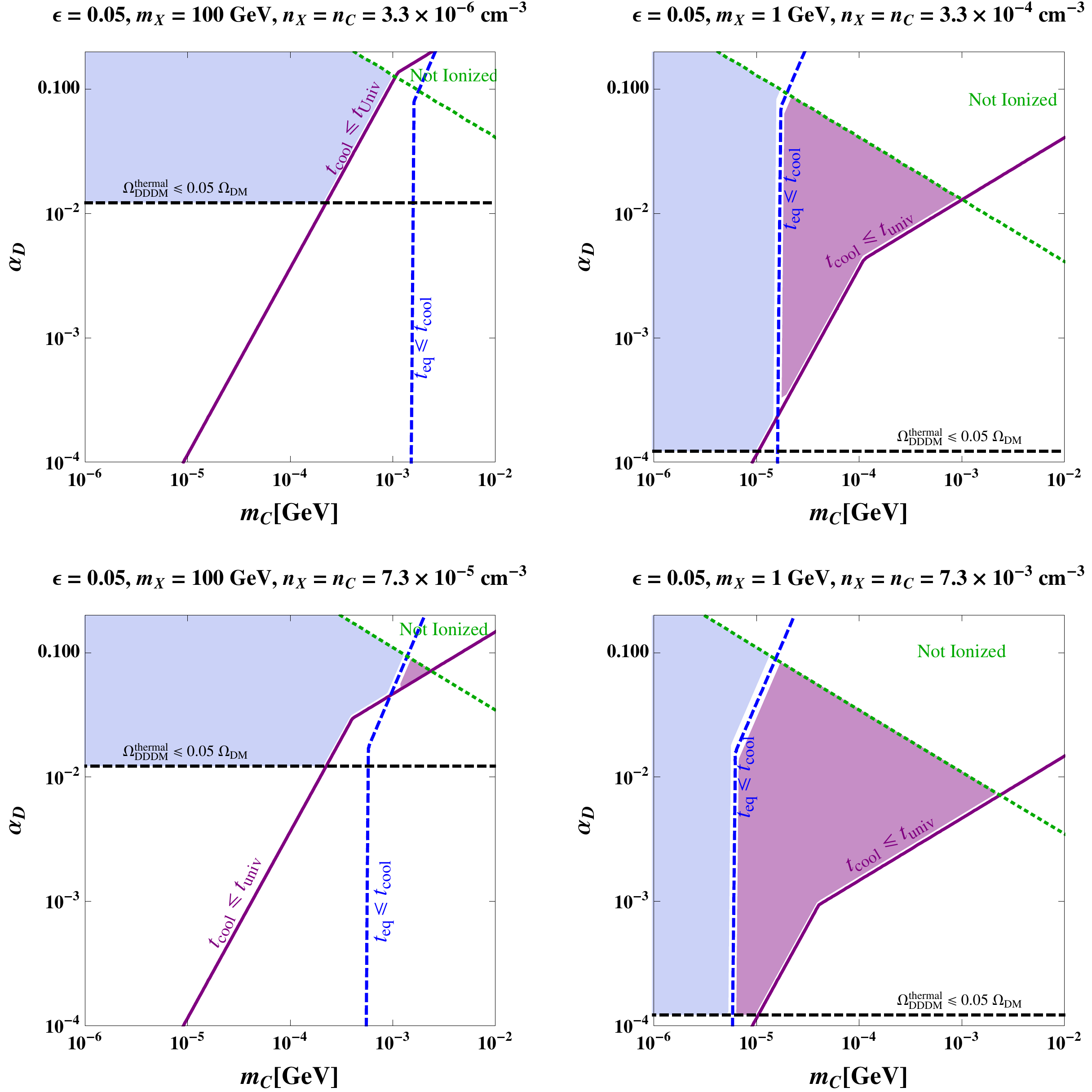}
\end{center}
\caption{Cooling in the $(m_C, \alpha_D)$ plane. The purple shaded region is the allowed region that cools adiabatically within the age of the universe. The light blue region cools, but with heavy and light particles out of equilibrium. We take redshift $z = 2$ and $T_D = T_{\rm CMB}/2$. The two plots on the left are for $m_X = 100$ GeV; on the right, $m_X = 1$ GeV. The upper plots are for a 110 kpc radius virial cluster; the lower plots, a 20 kpc NFW virial cluster. The solid purple curves show where the cooling time equals the age of the universe; they have a kink where Compton-dominated cooling (lower left) transitions to bremsstrahlung-dominated cooling (upper right). The dashed blue curve delineates fast equipartition of heavy and light particles. Below the dashed black curve, small $\alpha_D$ leads to a thermal relic $X,{\bar X}$ density in excess of the Oort limit. To the upper right of the dashed green curve, $B_{XC}$ is high enough that dark atoms are not ionized and bremsstrahlung and Compton cooling do not apply (but atomic processes might lead to cooling).}
\label{fig:rate}
\end{figure}%
%%%%%%%%%%%%%%%%%%%%%%%%%%%%%%%%%%%%%%%%%%%%%%%%%%%%%%

%%%%%%%%%%%%%%%%%%%%%%%%%%%%%%%%%%%%%%%%%%%%%%%%%%%%%%%
\begin{figure}[!h]\begin{center}
\includegraphics[width=0.4\textwidth]{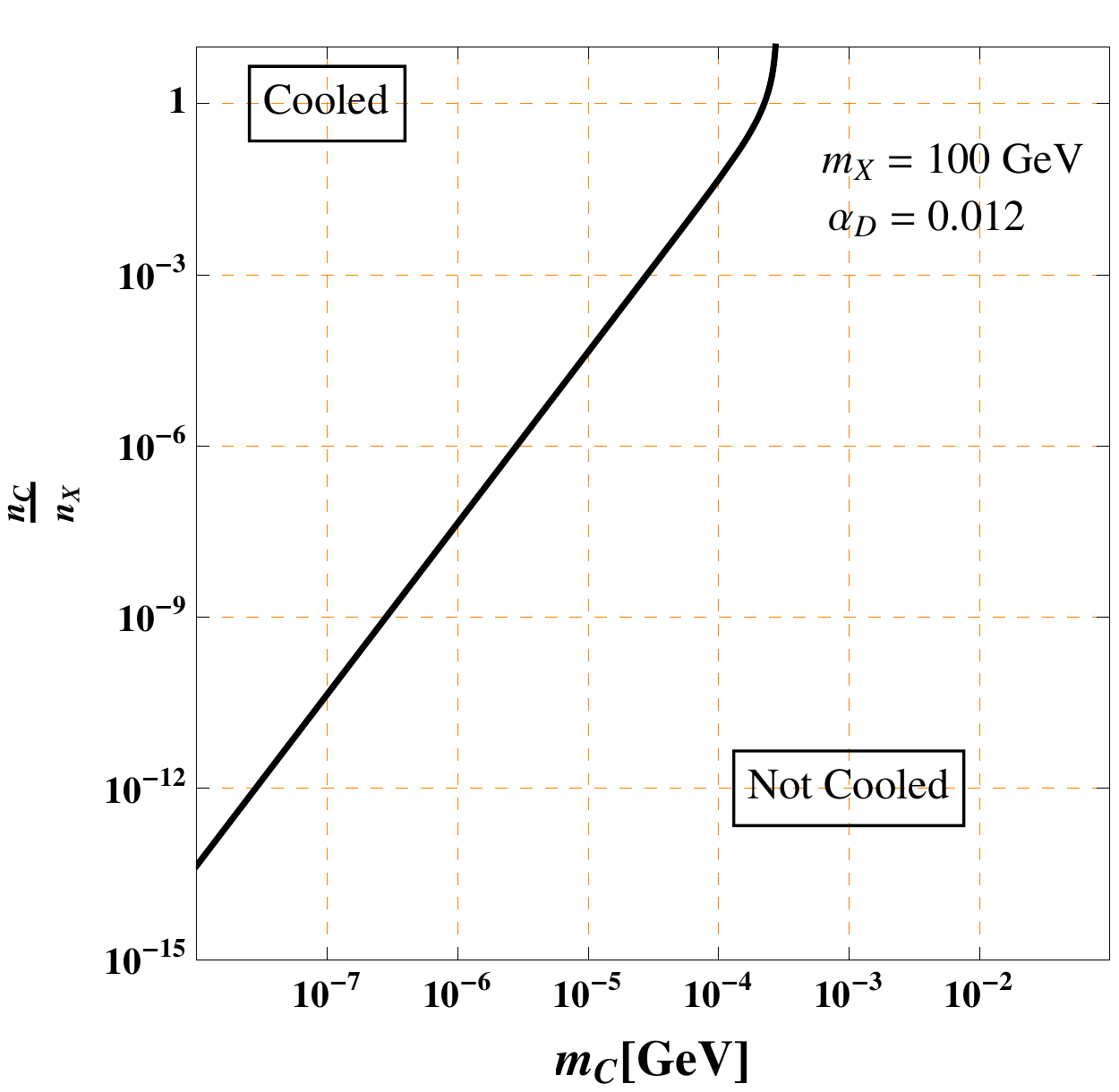} \quad \includegraphics[width=0.4\textwidth]{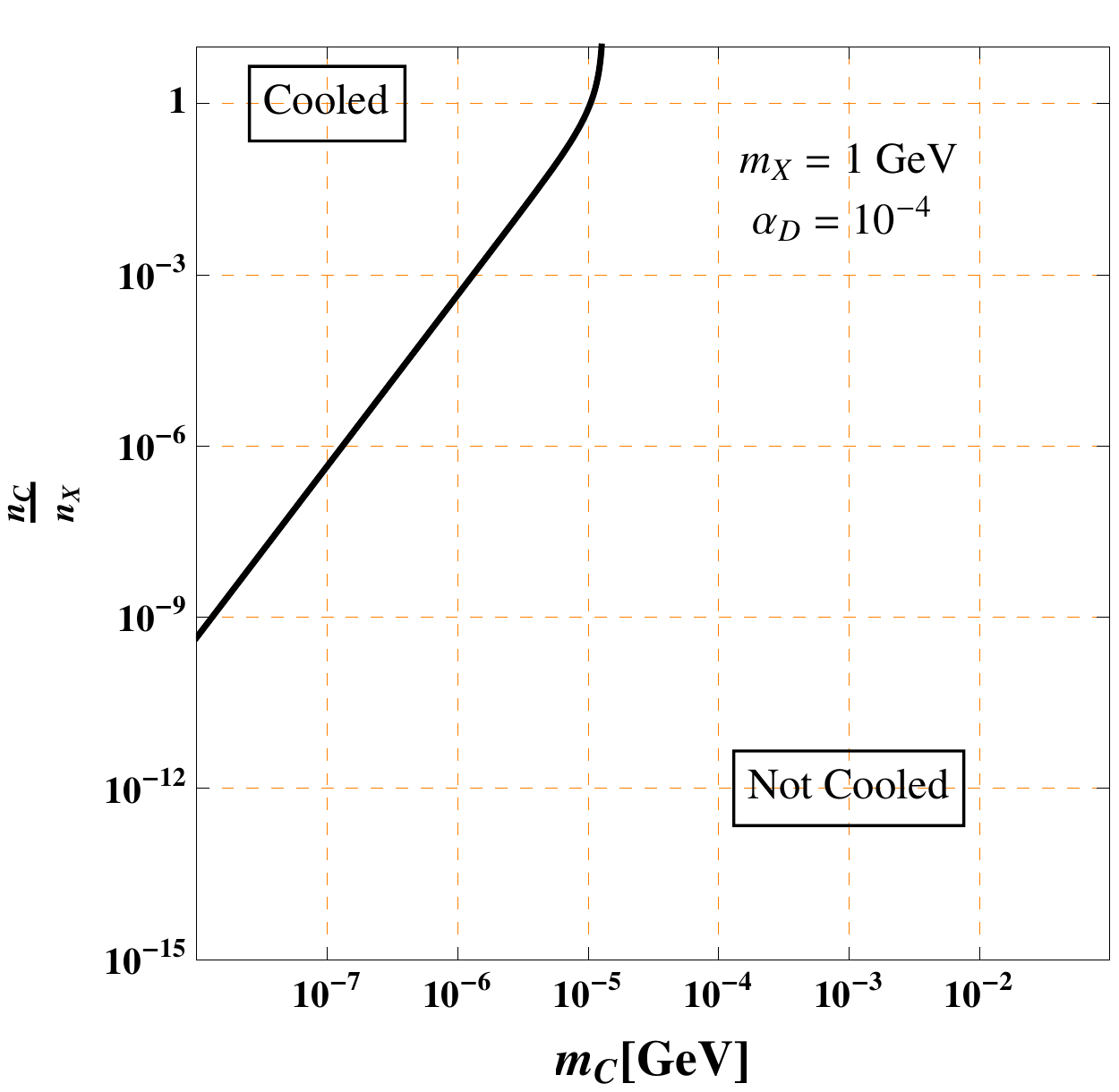}
\end{center}
\caption{Fixing $\alpha_D$ to yield the thermal relic abundance of DDDM as 5\% of the total DM relic abundance, the minimal $n_C/n_X$ with a cooling time scale equal to the age of the universe as a function of $m_C$ for $m_X = 100, 1$ GeV. We choose redshift to be $z=2$ and dark temperature $T_D = T_{\rm CMB}/2$. The bounds are from the Compton cooling process, which for the chosen $m_X$ and $\alpha_D$ dominates (so the bounds are independent of the DDDM density profile).}
\label{fig:nymin}
\end{figure}%
%%%%%%%%%%%%%%%%%%%%%%%%%%%%%%%%%%%%%%%%%%%%%%%%%%%%%%

In Fig~\ref{fig:rate} we present contours on which the cooling timescale is sufficiently rapid. We derive the bounds by assuming two different number densities $n_C = n_X$. First we assume conservatively that DDDM is uniformly distributed over a 110 kpc sphere as in Eq.~\ref{eq:numberdensity110kpc}. The bounds for $m_X = 100, 1$ GeV are shown in the upper row of Fig.~\ref{fig:rate}. Because DDDM will tend to fall into the halo, we expect that in fact cooling will be more rapid due to enhanced number density in the halo's central region. To obtain a more optimistic estimate, then, we estimate the time scale again using an NFW profile with a characteristic scale $R_s = 20$ kpc. We still use the virial theorem
\beq
\frac{1}{2}\frac{3G_NM(R_s)}{5R_s}=\frac{3}{2}T_{20}
\eeq
where $T_{20}$ is the temperature inside this region with radius $R_s$ = 20 kpc and $M(R_s)$ is the mass inside the region. Then the number density is chosen to be
\beq
n_C=n_X=\epsilon\frac{ 3M(R_s)}{4\pi R_s^3}\frac{1}{m_X}.
\eeq
The results are shown in the bottom row of Fig.~\ref{fig:rate}, and potentially allow masses an order of magnitude larger than the conservative estimate when bremsstrahlung is the dominant cooling mechanism. These plots show when efficient bremsstrahlung and/or Compton cooling can begin. Once cooling begins, increased density makes it more efficient, so the process will continue. For smaller $m_C$ and $\alpha_D$, Compton cooling could be faster than bremsstrahlung cooling. Thus the curves in Fig.~\ref{fig:rate} have kinks which correspond to transitions from bremsstrahlung cooling domination to Compton cooling domination.
We expect that both bremsstrahlung and Compton cooling will continue until heavy and light ions become cold enough to recombine into dark atoms.

From the left-hand plots in Fig.~\ref{fig:rate}, we see that for $m_X = 100$ GeV, there is a small region of parameter space where $m_C \approx 1$ MeV and $\alpha_D \approx 0.1$ where bremsstrahlung cooling happens within the age of the universe and the $X$ and $C$ particles are in equilibrium. A much larger region of parameter space with smaller $m_C$ has a fast cooling time but slow equipartition, so a better understanding of nonequilibrium cooling is needed to be certain of the fate of DDDM in this region. On the other hand, the right-hand plots of Fig.~\ref{fig:rate} show that a much larger region of parameter space cools adiabatically when $m_X = 1$ GeV. In particular, a Standard Model-like choice $m_X = 1$ GeV, $m_C \approx m_e$, and $\alpha_D \approx \alpha$ is on the edge of the region that cools efficiently by bremsstrahlung. (The SM gets some help from the baryon abundance being 15\% of all matter rather than $\epsilon = 0.05$ in our plot.) A larger region of parameter space down to $m_C = 10$ keV with $\alpha_D$ between 10$^{-4}$ and $10^{-1}$ can cool efficiently, typically through bremsstrahlung but with Compton cooling predominating at the smaller values of $\alpha_D$.

For the evaluations above, we assumed $n_C = n_X$, as is the case for fully asymmetric dark matter. This is not necessary. In Fig.~\ref{fig:nymin}, we fix $\alpha_D$ such that the thermal relic abundance of $X$ is 5\% of the total DM density (as in Figure~\ref{fig:relic}) and plot the minimal $n_C/n_X$ needed to have a cooling time scale shorter than the age of the universe. Combining both plots, we see that in most of the parameter space, sufficient cooling requires light DDDM with density greater than thermal (but comparable to the thermal abundance for $X$), which we will assume to be present. As discussed in Section~\ref{subsec:relicabundance}, we take the nonthermal component to be asymmetric.

Note also that in principle other cooling processes might occur. We expect atomic or even molecular processes or collisional cooling won't be important until below the expected recombination temperature, which is below the binding energy. However, note that it is only important at this point to establish that cooling can indeed occur, and that the temperature can be sufficiently low to form a thin disk no bigger than that for the baryons. Bremsstrahlung or Compton scattering with sufficiently light $C$ particles ensures this can indeed be the case.

Finally we would also like to emphasize that although we chose the fraction of DDDM to saturate the upper bound in our studies, the cosmological history remains similar and cooling and formation of a dark disk could still happen even if the fraction is smaller than 5\% of the total DM density. For example, if the heavy field has mass $m_X =$ 1 GeV, there is still parameter space in which the cooling time scale is shorter than the age of the Universe even if DDDM only constitutes 0.05\% of the total DM density. Hence the DDDM scenario will still survive even if the bound on DDDM relic abundance gets stronger.

%%%%%%%%%%%%%%%%%%%%%%%%%%%%%%%%%%%%%%%%
\section{Disk Formation}
\label{sec:diskform}
%%%%%%%%%%%%%%%%%%%%%%%%%%%%%%%%%%%%%%%%

Having established that DDDM can efficiently cool via bremsstrahlung or Compton scattering, we now consider how it will be distributed within the Milky Way. Like any matter falling into a halo, DDDM will have angular momentum, and so, as with baryonic matter, we expect that it will cool into a rotationally-supported disk. Because DDDM does not have supernova feedback and other processes that may be important in the evolution of the baryonic disk, this is not entirely obvious, and we rely on recent sophisticated numerical simulations in which disk formation occurs without including stellar and supernova feedback~\cite{Vogelsberger:2011hs,Torrey:2011qq}, rather than the earliest simulations in which baryons formed small clumps rather than disks~\cite{NavarroBenz91}. It is important to have further numerical work to confirm that this is true; DDDM that could form clumps instead of disks could also be extremely interesting.

We assume the disk mass distribution from Eq.~\ref{eq:distribution}. Assuming the disk scale length $R_d$ is much larger than the scale height $z_d$, we can neglect radial derivatives in the Jeans equation for an axisymmetric system and estimate, at the galactic center:
\beq
z_d  & \approx  & \sqrt{\frac{2 \overline{v_z^2}}{\pi G_N \rho_{\rm center}}} \approx \frac{16 \overline{v_z^2}R_d^2}{G_N \epsilon M^{\rm gal}_{\rm DM}} \nonumber \\
& \approx & 1.2\left(\frac{v^{\rm rms}_{z}}{10^{-3}}\right)^2 \frac{R_d}{\epsilon}.
\label{eq:scaleheight}
\eeq
where $v^{\rm rms}_z \equiv \sqrt{\overline{v_z^2}}$ is the velocity dispersion of DDDM in the vertical direction, $\rho_{\rm center} \equiv \rho(0,0)$ is the central mass density, and we have used Eq.~\ref{eq:distribution} in the second step. This estimate assumes the gravitational potential is dominated by the disk, i.e. it ignores the effects of baryons and of ordinary dark matter, but this is self-consistent to the extent that the disk is quite thin and thus the DDDM density is locally much larger than that of baryons and ordinary dark matter. We  estimate that the vertical velocity dispersion corresponds to the temperature at which cooling stops, $\overline{v_z^2} \approx T_{\rm cooled}/m_X$. We also assume that the disk scale length $R_d$ is comparable to that for baryons, around 3 kpc~\cite{BinneyTremaine}. Eq.~\ref{eq:scaleheight} should be viewed as a rough estimate; in particular, the detailed spatial distribution of DDDM may not precisely correspond to Eq.~\ref{eq:distribution}.

%%%%%%%%%%%%%%%%%%%%%%%%%%%%%%%%%%%%%%%%%%%%%%%%%%%%%%%
\begin{figure}[!h]\begin{center}
\includegraphics[width=0.45\textwidth]{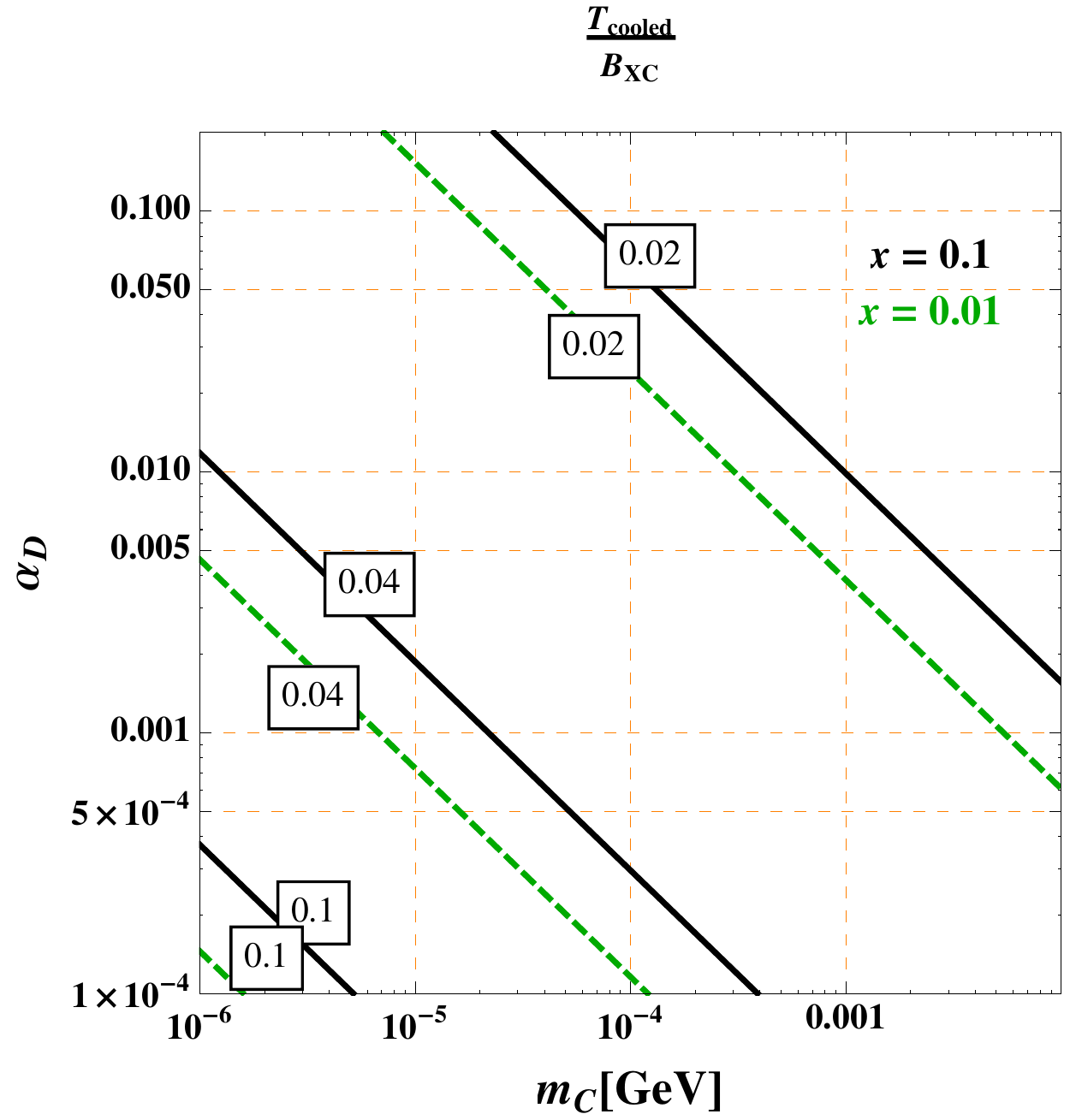}
\end{center}
\caption{Estimates of cooling temperature $T_{\rm cooled}/B_{XC}$ in the $(m_C, \alpha_d)$ plane. Black solid curves: ionization fraction $x$ = 0.1; green dashed curves: $x$= 0.01.}
\label{fig:Tcooled}
\end{figure}%
%%%%%%%%%%%%%%%%%%%%%%%%%%%%%%%%%%%%%%%%%%%%%%%%%%%%%%

In order to obtain concrete numbers from Eq.~\ref{eq:scaleheight}, we need an estimate of the final temperature $T_{\rm cooled}$. Bremsstrahlung and Compton cooling will cease to be efficient once the light particles are slow enough to recombine into dark $XC$ atoms, at temperatures low compared to the binding energy $B_{XC}$.\footnote{We are assuming an asymmetric scenario in which $X$ and $C$ are present, possibly with a symmetric X and ${\bar X}$ population, but ${\bar C}$ is absent. As discussed in Sec.~\ref{subsec:relicabundance}, a symmetric scenario may be possible in models with a more complex cosmology, in which case ${\bar X}{\bar C}$ recombination and $C{\bar C}$ annihilation could also occur during cooling.} A rough estimate of the temperature at which this occurs is found by solving the Saha equation,
\beq
\frac{n_X n_C}{n_{XC}n}=\frac{x^2}{1-x}=\frac{1}{n}\left(\frac{Tm_C}{2\pi}\right)^{3/2}\exp\left(-\frac{B_{XC}}{T}\right),
\label{eq:saha}
\eeq
where $n_{XC}$ is the bound state number density and $n=n_{XC}+n_X$ and $n_X=n_C$. The ionization fraction $x=n_C/n$. In order to obtain the relevant density, we assume that the gas has already cooled into a disk with scale radius $R_d$ and scale height $z_d$
\beq
n=\frac{\rho_0}{m_X}=\frac{\epsilon M_{\rm gal}}{8 \pi R_d^2z_d m_X}=\frac{G_N (\epsilon M_{\rm gal})^2}{128\pi R_d^4 T}.
\label{eq:numberden}
\eeq
Combining Eq.~\ref{eq:saha} and Eq.~\ref{eq:numberden} and requiring the ionization fraction to be smaller than 1, e.g., 0.1 or 0.01, we find the results shown in Figure~\ref{fig:Tcooled}, which we summarize as:
\beq
T_{\rm cooled} \sim (0.02 - 0.2) B_{XC}.
\eeq
Thus, we expect that cooling stops at a temperature of about 10\% of the binding energy. This leads to estimates of the disk scale height that are substantially thinner than the baryonic disk. Over the bulk of parameter space, we find that the results are well-described by a power law:
\beq
z_d \approx 2.5\,{\rm pc}\left(\frac{\alpha_D}{0.02}\right)^{2} \frac{m_C}{10^{-3}~{\rm GeV}}\frac{100~{\rm GeV}}{m_X}
\label{eq:zdestimate}
\eeq
The $1/m_X$ scaling arises because, at a given temperature, the velocity of the dark atoms is smaller at larger $X$ masses. In other words, we expect large boost factors for weak-scale dark matter because it is much heavier than baryons. For lighter $X$ particles, smaller values of $\alpha_D$ can still allow cooling and a thermal abundance of the symmetric component, in which case we again can get large boost factors simply because the binding energy, and hence the temperature of dark atoms relative to baryonic atoms, can be much smaller.

Further cooling could occur, as in the baryon sector, through molecular processes; on the other hand, heating processes could also occur that would thicken the disk. For instance, the gravitational influence of interstellar clouds on the vertical distribution of stars in the Milky Way is important~\cite{SpitzerSchwarzschild1,SpitzerSchwarzschild2}. A molecular cloud that accelerates stars will also accelerate gas particles, like the $XC$ bound states, and this could thicken the disk. However, stars are collisionless, while $XC$ bound states could cool down again, so we expect the velocity dispersion imparted by interstellar clouds to be smaller for DDDM than for stars. In the absence of a more thorough treatment of such possible heating mechanisms, we can only say that the true disk thickness is expected to lie between Eq.~\ref{eq:zdestimate} and the height of the baryonic gaseous disk (on the order of 100 parsecs). It would be very interesting to see if simulations could provide a more robust estimate of the disk height, which is crucial for understanding the possible enhancement in dark matter detection signals.

The angle between the baryon and DDDM disks also plays a key role in the observability of DDDM, especially for direct and indirect detection. We would expect that gravity would tend to align these structures in a timescale set, very approximately, by $t \sim R\sqrt{R/GM_{\rm disk}} \sim 10^7~{\rm yr}$. In fact, even as the galaxy first formed, the angular momentum vectors of baryons and DDDM could already have been approximately aligned, because filaments in the cosmic web define preferred directions for accretion. Recent numerical simulations of the galaxy~\cite{Hahn:2010ma} have found that the stellar and gaseous components of the baryonic disk are typically aligned to within about 7$^\circ$, and the angular momentum vector of dark matter in the inner halo is somewhat less aligned, with a median angle of 18$^\circ$ to the angular momentum of the gaseous disk. The fact that simulations see a much better alignment of the angular momentum of the baryonic disk with the angular momentum of dark matter in the {\em inner part} of the halo~\cite{Bett:2009rn,Hahn:2010ma}, rather than the entire halo, is reflective of the gravitational alignment that we expect to happen between the two disks in our model. Because we expect approximate alignment, indirect detection signals from the galactic plane might be expected as we discuss in the following section.

Dark disks may also arise from ordinary dark matter accreting onto the stellar disk~\cite{Read:2008fh, Bruch:2008rx, Bruch:2009rp}. Their phenomenology of direct detection and solar capture are similar to what we will discuss in Sec.~\ref{sec:detection}, but our mechanism to generate the disk is completely different. If such a dark disk of ordinary cold dark matter exists as well, it will be aligned with the baryonic disk and its effect on the DDDM disk will be similar to that of baryons. Interestingly, the dynamics of accretion might also add ordinary DM to the DDDM disk, if it is not aligned with the baryonic disk. Again, detailed simulations are needed to quantify the effects of these various disks on each other.

Finally we want to comment on the ``thin'' and ``thick'' disks. It is known that in the Milky Way's stellar disk, different stellar sub-populations have different vertical scale heights, their thickness increasing with age. But as argued by~\cite{Bovy:2011zx}, there is no ``thick disk'' that is characterized as a seperate component. For our DDDM scenario, it is unclear whether compact objects such as ``dark stars'' would form, at least in the simplest $U(1)$ model given the lack of nuclear reactions, and even less clear is the effect of compact dark objects on the disk height.

%%%%%%%%%%%%%%%%%%%%%%%%%%%%%%%%%%%%%%%%
\section{Indirect and Direct Detection}
\label{sec:detection}
%%%%%%%%%%%%%%%%%%%%%%%%%%%%%%%%%%%%%%%%

\subsection{Indirect detection}

\begin{figure}[!h]\begin{center}
\includegraphics[width=0.45\textwidth]{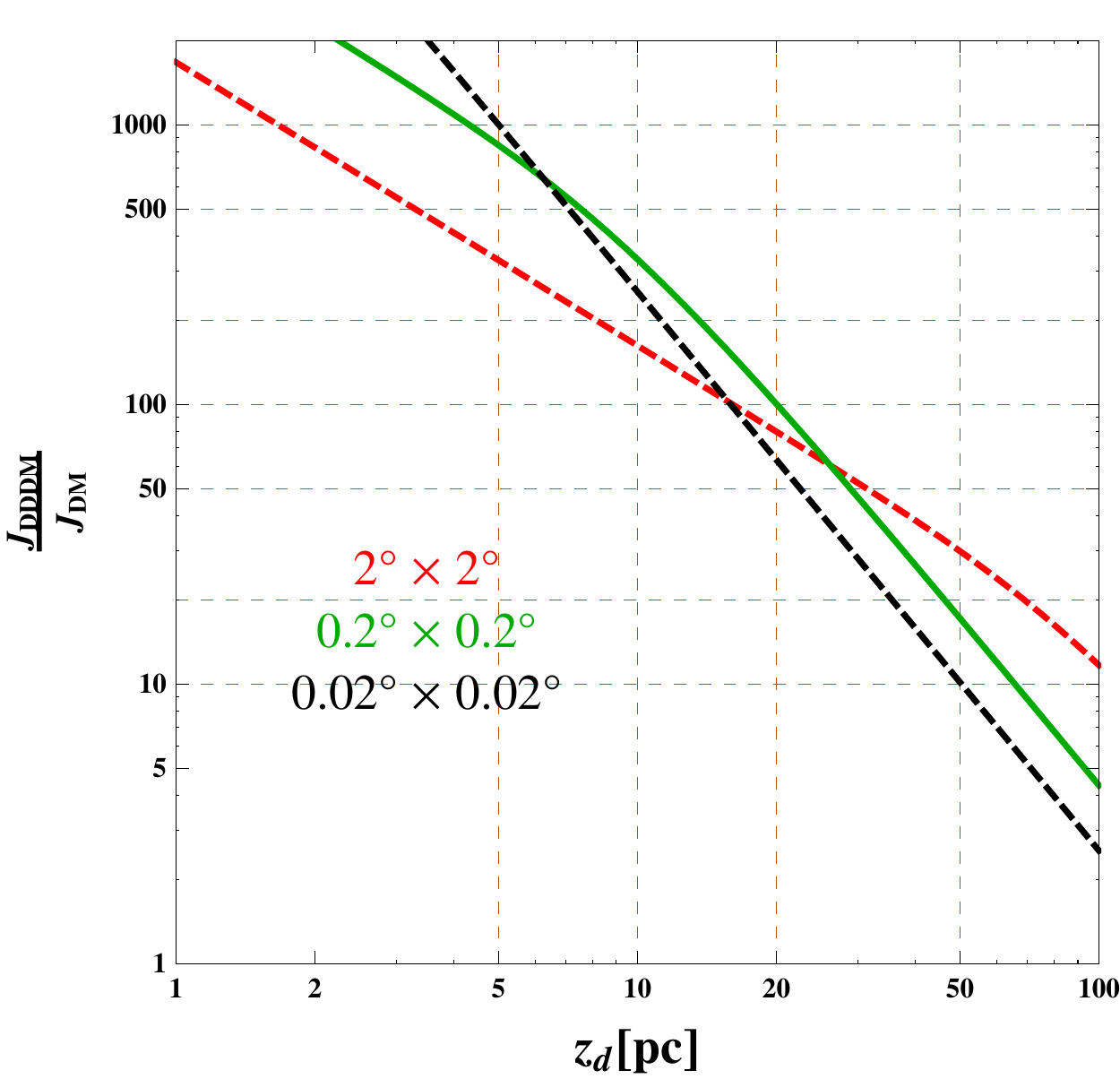}
\end{center}
\caption{Local density enhancement in DDDM, as a function of disk scale height $z_d$, in a square region around the GC fixing $\epsilon = 0.05$ that DDDM is 5\% of the total DM density. Red: region within $b \subset (-1^\circ, 1^\circ), l \subset (-1^\circ, 1^\circ)$. Green: region within $b \subset (-0.1^\circ, 0.1^\circ), l \subset (-0.1^\circ, 0.1^\circ)$ (current Fermi-LAT angular resolution). Black: region within $b \subset (-0.01^\circ, 0.01^\circ), l \subset (-0.01^\circ, 0.01^\circ)$.  }
\label{fig:indirect}
\end{figure}

A dramatic signal of DDDM can arise from annihilation of dark matter particles with their antiparticles, e.g. annihilation of residual ionized $X$ with ${\bar X}$ into gamma rays for example. Because photons travel unimpeded to us, such a signal could in principle provide a map of the dark disk on the sky, giving striking visual confiFrFFrmation that dark matter has cooled into a structure distinct from a typical halo. The gamma-ray intensity in a given direction is the line-of-sight integral of the DM number density squared along a given direction,
 \beq
 \frac{d \Phi_\gamma}{d E_\gamma}=\frac{1}{8\pi}\frac{\langle \sigma v \rangle_{\gamma\gamma}}{m_{DM}^2} 2 \delta(E-E_\gamma)d_\odot \rho_\odot^2 J,
 \eeq
 with:
 \beq
 J=\int_{\rm roi} db \, dl \int_{\rm l.o.s} \frac{ds}{d_\odot}\cos b \left(\frac{\rho(r)}{\rho_\odot}\right)^2,
 \eeq
 where $\rho_\odot$ is the normal DM density at the Sun, $\rho_\odot=0.3$ GeV cm$^{-3}$. $d_\odot$ is the distance from the Sun to the galactic center (GC), $d_\odot \approx 8.3$ kpc. The integral is over the region of interest (roi) at the GC. The smallest region centered around the GC that the Fermi-LAT experiment is sensitive to is a $0.2^\circ \times 0.2^\circ$ square due to finite angular resolution, which corresponds to a 28 pc $\times$ 28 pc region around the GC. Thus for a disk height $z_d > 28$ pc, we expect that $J_{\rm DDDM}$ scales as $z_d^{-2}$. Fig.~\ref{fig:indirect} shows the local density enhancement of DDDM compared to the normal DM defined as
 \beq
 \frac{J_{\rm DDDM}}{J_{\rm DM} }\nonumber
 \eeq
 as a function of the DDDM disk height $z_d$, where for normal DM, we used an Einasto profile
 \beq
 \rho_{\rm Einasto}(r)=\rho_s \exp{\left(-(2/\alpha_E)\left((r/r_s)^{\alpha_E}-1\right)\right)},
 \eeq
 with $r_s$ = 20 kpc and $\alpha_E = 0.17$. $\rho_s$ is fixed to achieve the correct $\rho_\odot$. The resulting boost factor arises not only from the compression of the disk in the vertical direction, but also because the disk scale length in the radial direction is somewhat smaller than the radial spread in the distribution of ordinary dark matter.

%%%%%%%%%%%%%%%%%%%%%%%%%%%%%%%%%%%%%%%%%%%%%%%%%%%%%%%
%%%%%%%%%%%%%%%%%%%%%%%%%%%%%%%%%%%%%%%%%%%%%%%%%%%%%%

%%%%%%%%%%%%%%%%%%%%%%%%%%%%%%%%%%%%%%%%%%%%%%%%%%%%%%%
\begin{figure}[!h]\begin{center}
\includegraphics[width=0.5\textwidth]{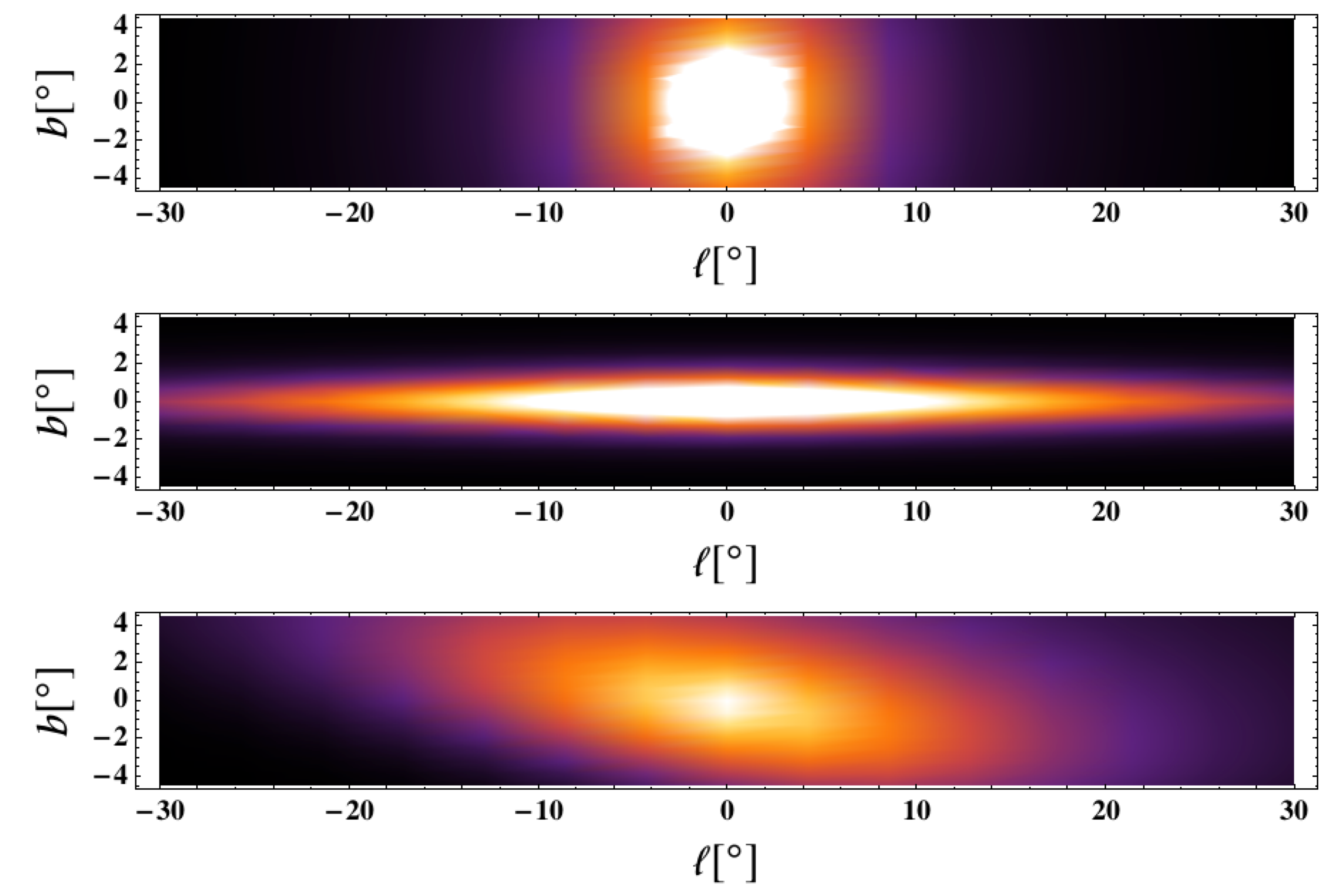}
\end{center}
\caption{Sky maps of the photon flux shape in arbitrary units for different DM profiles. Upper: Normal DM with an Einasto profile. Middle: DDDM in a disk aligned with our disk. Lower:  DDDM in a disk misaligned with our disk by 18$^\circ$. The DDDM images have disk scale height $z_d = 100$ pc.}
\label{fig:skymaps}
\end{figure}%
%%%%%%%%%%%%%%%%%%%%%%%%%%%%%%%%%%%%%%%%%%%%%%%%%%%%%%

Clearly an enhanced DDDM density would be distinguishable. Even if the density enhancement $J_{\rm DDDM}/J_{\rm DM}$ integrated over the region of interest is modest, the distribution of photons within this region---and especially at larger distances from the galactic center---can be radically different for DDDM and ordinary dark matter. Some illustrations of the photon flux over the sky are shown in Figure~\ref{fig:skymaps}.

Another feature of a possible indirect detection signal from DDDM annihilation is that a larger Sommerfeld enhancement can arise due to the smaller velocity dispersion of DDDM. It is usually assumed that the DM halo is approximately isothermal, and thus the velocity distribution is mostly Gaussian with a dispersion $\sim 10^{-3} c$. However, DDDM travels in circular orbits around the GC and its velocity dispersion could be much smaller than $10^{-3} c$. Specifically, its velocity dispersion is determined by $T_{\rm cooled}$ through $\overline{v^2} = 3 T_{\rm cooled}/m_X$. For example, in the parameter space where bremsstrahlung or Compton cooling is efficient as shown in Sec.~\ref{sec:cooling}, $\overline{v_z^2} < 10^{-9}$ for $m_X = 100$ GeV. Thus the Sommerfeld enhancement factor, which scales as $1/v$, could be increased by a factor of 10 or more compared to non-dissipative DM charged under U(1)$_D$ with the same charge and mass.

In summary, for indirect detection, the DDDM scenario could easily accommodate a large boost factor from local density enhancement in the range (10--1000) depending on the disk height. Also due to a smaller velocity dispersion, DDDM could have a larger Sommerfeld enhancement, giving rise to another boost factor of ${\cal O} (100)$. Thus the DDDM scenario could easily explain the suggestion of a Fermi photon line at around 130 GeV without large couplings or tuned masses. Again, we would like to emphasize that this photon line only serves as an example that the DDDM scenario could lead to interesting and distinctive indirect detection signals.

Throughout this discussion we have assumed that the DDDM disk and the ordinary dark matter are centered on the same location. The Fermi 130 GeV line is arguably off center~\cite{Tempel:2012ey,Su:2012ft}, which has provoked some debate, with numerical simulations showing that dark matter may be displaced from the galactic center~\cite{Kuhlen:2012qw} and others arguing that tidal disruption prohibits such a displacement~\cite{Gorbunov:2012sk}. It would be interesting to explore the similar question of whether the DDDM disk and the baryonic disk can be centered on different locations.

\subsection{Direct detection}

Direct detection of dark matter could in principle be possible if the Earth is located within the DDDM disk. In the most optimistic case when the Sun is in the DDDM disk, the DDDM density at the position of the Sun could be as large as 6 GeV/cm$^3$, 20 times as large as the normal DM density, for $\epsilon = 0.05$ and $z_d = 100$ pc. However, the spectrum of DDDM scattering off nucleons would be very different from that of an ordinary WIMP with the same mass, and we will see that the kinetic energy of DDDM is too low to produce a measurable signal in conventional direct detection experiments.

The kinematics of direct detection involves a dark matter particle moving with a nonrelativistic velocity $v$ in the lab frame, which scatters off a stationary nucleus. Depending on the scattering angle, the recoil energy imparted to the nucleus can take any value between zero and
\beq
E_R^{\rm max}& = & \frac{2 \mu_N^2}{m_N} v_X^2  \label{eq:recoil}\\
& \approx & 0.5\, {\rm keVnr}\left( \frac{\mu_N}{50\,{\rm GeV}} \right)^2\frac{100\,{\rm GeV}}{m_N}\left( \frac{v_X}{10^{-4}}\right)^2,\nonumber
\eeq
where $v_X$ is the dark matter velocity, $m_N$ is the mass of the target atom, and $\mu_N$ is the reduced mass of the DDDM--nucleus system. Most experiments are sensitive to energies above a threshold value of $E_R$, below which noise and various backgrounds can overwhelm any possible dark matter signal. Having a threshold $E_R$ corresponds to being sensitive to a minimum value of the dark matter velocity $v_X$. For ordinary dark matter, there is a broad spectrum of velocities that can be approximately modeled as an isothermal distribution with typical velocity $10^{-3} c$. But DDDM is not ordinary dark matter: after cooling, it is in the form of a rotationally supported disk, and a typical DDDM particle will move in a circular orbit around the center of the galaxy. Near the Sun, both DDDM and the solar system would be in approximately the same circular orbit, so the large radial component of their velocity will be identical. Only deviations from this typical circular velocity can contribute to scattering. The rate for spin-independent elastic scattering is:
\beq
\frac{d \Gamma}{d E_R}=N_t \frac{m_N \rho_X \sigma_n}{2 m_X \mu_{n}^2 v_X} A^2 F(E_R)^2{E_R}\theta(v_X-v_{\rm min}),
\eeq
where $N_t,~m_N$ and $A$ are the number, mass, and atomic number of the target atoms; $m_X$, $\rho_X$ and $v_X$ are the the mass, local density, and velocity of DDDM at the Sun; $\sigma_n$ is the zero-momentum spin-independent DDDM--nucleon scattering cross section; $\mu_{n}$ is the reduced mass of the DDDM--{\em nucleon} system; $F(E_R)^2$ is the nuclear form factor; and $v_{\rm min}$ the minimum DDDM velocity needed to create a recoil with recoil energy $E_R$. Before taking into account the nuclear form factor $F(E_R)^2$, the spectrum would be flat between 0 and $E_R^{\rm max}$. However, the nuclear form factor $F^2(E_R)$ is in general an exponentially falling function, which suppresses higher energy recoils, yielding a falling spectrum with an end point at $E_R^{\rm max}$.

The typical threshold for current direct detection experiments is a few keV in nuclear recoil (e.g., the CoGeNT threshold is 0.5 keVee $\sim 2$ keVnr). Equation~\ref{eq:recoil} shows that the velocities in the lab frame need to be larger than $10^{-4} c$ in order for DDDM to produce hard enough recoils to be detected at such experiments. There are several sources of relative velocity between the detector and the DDDM. One is the peculiar velocity of the Sun, which does not have a perfectly circular orbit. Another is the motion of the Earth around the Sun. Both of these velocities are around $10^{-4} c$, too small to give an easily detectable signal. Other sources of relative velocity could arise from inhomogeneities in the disk that lead to deviations from perfectly circular orbits. For instance, the spiral arms of the baryonic disk are density waves, analogous to traffic jams, across which the radial velocities of stars vary. The Sun is in such a spiral arm, and the DDDM disk may also have density waves or other structures in which velocities differ. However, given that the spiral arm in our disk only modifies stellar velocities by $\sim 10^{-4} c$, it is not clear that such effects can be large enough to change our conclusion.

In general, it is interesting that due to the DDDM's small velocity, only the energy bins close to an experimental threshold could be sensitive to DDDM scattering. So far, the importance of energy calibration around the threshold has been mostly emphasized for ruling in or out the light DM scenario. Yet from the discussions above, pushing the thresholds of direct direction lower could also be important for the DDDM scenario, or in general, DM with a small velocity~\cite{Bruch:2008rx}.

Another possible way to detect DDDM directly is to look for single- or few-electron events if it scatters with electrons, causing ionization of atoms in a detector target material. In particular, dual-phase liquid xenon detectors could have sensitivity to such small ionization signals~\cite{Edwards:2007, Santos:2011, Aprile:2001}.

The low relative velocity would tend to suppress direct detection even in the most optimistic case, when the Earth is directly inside the DDDM disk. Another possible suppression mechanism is that the Earth could be located in a region of low DDDM density. The Earth sits about 10 to 20 parsecs above the galactic plane~\cite{Joshi:2007ww} and about 8 kpc from the galactic center~\cite{Ghez:2008ms, Gillessen:2008qv}. A 5$^\circ$ inclination of a $z_d = 100$ pc DDDM disk would suppress the local DDDM density by a factor of 10 compared to the normal DM density at the Sun, $\rho = 0.3$ GeV/cm$^3$. Hence, even improved low-threshold experiments could not completely rule out DDDM, since it is always possible that the density near the Earth is simply too small to observe.

\subsection{Solar capture}

Another possibility for detection arises from solar capture. As dark matter particles pass through the Sun, they could scatter off nuclei inside the Sun and become gravitationally bound. With subsequent scattering (between themselves and nuclei), they could eventually accumulate in the center of the Sun. Captured $X$ and ${\bar X}$ particles could subsequently annihilate into various SM final states. For instance, they could annihilate into $Z\gamma$ or $ZZ$ through the same loop of charged particles that leads to monochromatic photon lines. $Z$s would subsequently decay into energetic neutrinos, which could be observed in neutrino telescopes on Earth, such as IceCube.

Currently IceCube constrains the spin-independent capture rate to be $C_{\rm SI}^{\odot}\sim10^{22}$ s$^{-1}$ for a 100 GeV DM particle, which for an ordinary DM particle corresponds to a constraint on the DM--nucleon cross section of $\sigma_{SI}^p \sim 6.0 \times 10^{-43}$ cm$^2$~\cite{Aartsen:2012kia}. This interpretation of the data relies on the assumption that DM has come into equilibrium in the Sun, at which point the capture and annihilation rates are comparable; see e.g. the review~\cite{Jungman:1995df}. If the equilibrium is not achieved, the DM annihilation rate would be suppressed compared to the capture rate.

In the DDDM scenario, two factors could enhance the solar capture rate (given the same DM--nucleon cross section). First, if the sun is inside the dark disk, DDDM should have a larger local density, $\sim 10$ times as large as the ordinary DM local density near the sun. The capture rate could also be enhanced by a larger gravitational Sommerfeld enhancement scaling as the inverse of the velocity dispersion~\cite{Gould:1987ir}, again a factor of 10 compared to that of ordinary DM. This enhancement due to lower velocity also happens in a dark disk made of ordinary dark matter~\cite{Bruch:2009rp}. The DDDM annihilation rate is also Sommerfeld enhanced due to the long-range U(1)$_D$. Thus we expect that DDDM always comes into equilibrium in the Sun unless the DM--nucleon cross section is very small. The DDDM--nucleon cross section varies from about $10^{-49}$ cm$^2$~\cite{Weiner:2012cb}, if DDDM interacts only through a loop-suppressed coupling to SM photons, to $10^{-44}$ cm$^2$ if DDDM couples to SM gluons at the one-loop order. This is equivalent to a capture rate in the range $10^{17}$ s$^{-1}$--$10^{22}$ s$^{-1}$, assuming that the DDDM capture rate is 100 times as large as that of ordinary DM with the same cross section for scattering on nucleons. Thus DDDM could potentially lead to a signal in the ongoing IceCube experiment. It is interesting that, although small velocities make direct detection more difficult, they enhance the solar capture rate and could lead to larger signals at IceCube, which then plays an important complementary role.

Though it is not possible in our minimal models, more complicated models could also lead to possible signals from high energy gamma rays or charged particles such as $e^\pm$ near the Sun. For instance, DDDM could annihilate into metastable intermediate particles that decay outside the Sun into photons or $e^+, e^-$ pairs; or if DDDM particles scatter inelastically, captured DDDM can be bound in elliptical orbits of order the size of the Sun and can then annihilate outside of the Sun~\cite{Schuster:2009fc}.

%%%%%%%%%%%%%%%%%%%%%%%%%%%%%%%%%%%%%%%%%%%%
\subsection{LHC Searches}
%%%%%%%%%%%%%%%%%%%%%%%%%%%%%%%%%%%%%

In principle, the LHC can search for dark matter as well. WIMP searches can proceed if the WIMP is part of a larger sector, such as a supersymmetric theory where charged superpartners can decay to the LSP. Such searches are unlikely to apply for this new sector unless it is also part of a BSM model, which we leave an open question.

Other searches~\cite{Birkedal:2004xn,Goodman:2010yf,Bai:2010hh} rely on crossing the interaction (see Appendix~\ref{app:models}) responsible for either direct or indirect searches. Ref.~\cite{Weiner:2012cb} studies whether an operator that produces the Fermi signal can also lead to a detectable LHC signal. They concluded that the signal is barely detectable when there is no large boost factor. Our large boost factor implies a smaller strength matrix element that will not be observable in the near future.

%%%%%%%%%%%%%%%%%%%%%%%%%%%%%%%%%%%%%%%%
\section{Conclusions and Future Work}
\label{sec:conclusions}
%%%%%%%%%%%%%%%%%%%%%%%%%%%%%%%%%%%%%%%%

In this paper we have shown that it is possible for a subdominant component with up to 15\% of all dark matter and dissipative dynamics to collapse into a disk similar to the baryonic one. In effect, such matter behaves much like a new kind of ordinary matter, constituting a hidden world neighboring our own. If it annihilates to visible-sector particles, we could see a striking enhanced indirect detection signal distributed on the sky very differently from that expected for ordinary dark matter. Even without such indirect detection channels, it is very possible that such new forms of matter could be detected through their gravitational interactions with other matter. We have given some simple estimates of the properties of Double-Disk Dark Matter, but much remains to be done. Here we will briefly outline some important directions for future work.

{\bf Numerical simulations.} Questions such as small-scale structure and the expected alignment of the DDDM and baryonic disks would best be answered through numerical simulations of galaxy formation (e.g. in mixed $N$-body/hydrodynamical codes). A fully correct picture, especially in the case that equipartition between heavy and light particles is not fast, would likely require modeling dark electric and magnetic fields as well. Such numerical simulations might also shed light on the expected velocity distribution of DDDM near the Sun, which is important for understanding whether direct detection could ultimately be possible.

{\bf Large-scale structure.} DDDM could impact the large-scale structure of the universe in ways that might be detectable in the CMB, galaxy/galaxy correlation functions, or other observables. For instance, dark acoustic oscillations are a possible signal~\cite{CyrRacine:2012fz}. It would be interesting to determine whether surveys of large-scale structure, possibly including upcoming 21 cm observations that probe the cosmic dark ages, could be sensitive to the existence of DDDM.

{\bf Small-scale structure.} $\Lambda$CDM simulations have problems at small scales such as overly large cusp predictions and too many satellites. The potential for DDDM to address these problems deserves study. After the disk forms and cools, regions within the gas of dark atoms can suffer gravitational collapse, perhaps leading to interesting small-scale structure and compact objects.

{\bf Chemistry and nuclear physics.} It will also be interesting to study dark matter chemistry, which should resemble hydrogen chemistry, as well as nuclear physics if additional interactions are included.  Further cooling processes that depend on dark atoms and molecules will be interesting to study in this case.

{\bf Observational tests for a disk.} The Gaia satellite, or other surveys of stars in the Milky Way, will study star velocities with unprecedented precision. It is important to see how this can be used to map out the distribution of dark matter and test for the presence of structures like a DDDM disk through their gravitational effects. Other possible tests could be microlensing from compact DDDM objects or lensing from the net effect of the DDDM disk on light from distant objects. From the particle physics perspective, models in which DDDM exists but can be only be detected gravitationally are conceivable, so it is vital to understand whether current or future observations can directly probe its gravitational effects.

{\bf Indirect detection bounds.} New analyses of existing data, for instance from Fermi-LAT, could be used to set limits on the annihilation rate of DDDM into Standard Model particles. These analyses would differ from the standard analysis because DDDM's spatial distribution is very different from a typical halo profile.

{\bf Nonthermal cosmologies.} It will be interesting to explore models that generate the dark sector asymmetry and possibly relate it to the baryon asymmetry. A more thorough exploration of possible scenarios generating a nonthermal symmetric $C, {\bar C}$ component at late times could also be interesting.

The various issues highlighted above may not all be decoupled. For example, it has been proposed that the cusp/core problem is related to supernova explosions that flatten the dark matter cusp into a core. This is in tension with the low star-formation efficiency suggested by the missing satellites problem~\cite{Penarrubia:2012bb}. But suppose that the DDDM sector involves violent small scale events that, like supernovae, inject energy that could flatten out the cusp. This may not happen in our minimal model, but is conceivable in a DDDM model with a closer resemblance to the Standard Model, for instance. Such events may not be observable in visible light, and so the tension with the missing satellites problem may not exist in this case. This is just one of many possible directions that could tie together the physics and astrophysics of DDDM in novel ways.

We emphasize that, although a dizzying array of particle physics models have been proposed for dark matter, most appear from the astrophysical perspective as the same cold, collisionless dark matter. Explorations of dark matter with different astrophysical consequences are mostly limited to warm dark matter and self-interacting dark matter, with the latter usually assuming a fixed cross section for point-like interactions or, at times, velocity dependence~\cite{Feng:2009hw,Buckley:2009in,Loeb:2010gj,Vogelsberger:2012ku,Tulin:2012wi,Tulin:2013teo}; though see also~\cite{SteinhardtNewPaper}.  Such scenarios with two components of dark matter---the dominant one essentially noninteracting and a small component with self interactions---introduce many new possibilities and are as yet only weakly constrained. We believe that Partially Interacting Dark Matter and, in particular, Double-Disk Dark Matter go far beyond  standard scenarios for dark matter and offer very novel prospects for dark matter astrophysics. The phenomena we have discussed in this paper could be just the tip of the iceberg.

\appendix
\section{Satisfying Kinetic Mixing Constraints in  Models of DDDM }
\label{app:models}

\subsection{Kinetic Mixing Constraints}

Ordinary photons and dark photons can mix through the operator $\kappa F_{\mu \nu}F_D^{\mu \nu}$~\cite{Holdom:1985ag}. Assuming (as we have throughout the paper) that dark photons are exactly massless, such a mixing means that ordinary matter has a small charge under U(1)$_D$ and dark charged matter has a small charge under U(1)$_{\rm EM}$. This would keep the DDDM sector and visible sector coupled to dangerously low temperatures. In particular,  in the early universe, it would allow the Standard Model plasma to produce $C$ and ${\bar C}$ through interactions like $e^+ e^- \to C {\bar C}$ or $\gamma \to C {\bar C}$. When $m_C \simlt 1$ MeV this will overproduce relativistic $C$ particles and violate the bound from BBN discussed in Sec.~\ref{subsec:temp}, unless~\cite{Davidson:2000hf}
\beq
\kappa \lesssim 10^{-9}.
\label{eq:kappabound}
\eeq
For even lighter $C$ masses, $m_C < 10~{\rm keV}$, there is an even stronger bound of
$\kappa\lesssim 10^{-13}~\rm{to}~10^{-14} $ from the cooling of red giants and white dwarfs (for details see~\cite{Davidson:2000hf} and references therein).

\begin{figure}[!h]
\centering
\includegraphics[width=0.25\textwidth]{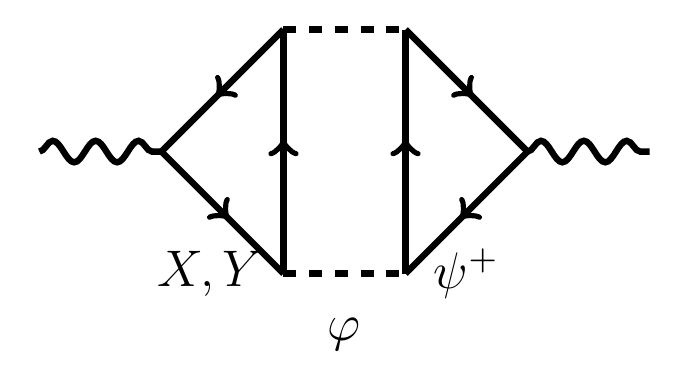}
\caption{A possible 3-loop contribution to kinetic mixing when no particles are charged under both sectors, which is zero because it involves both particles and antiparticles.}
\label{fig:kineticmix3loop}
\end{figure}

Kinetic mixing has the potential to be dangerous because, as a marginal operator, it can be generated at any scale. Even GUT-scale particles charged under both groups could lead to violation of the experimental bound. Despite this danger there are several ways of avoiding the bound. First, we observe that it is consistent to simply set the kinetic mixing to zero, if there are no particles at any scale charged under both U(1)$_Y$ and U(1)$_D$. This is true even if particles charged under the two groups interact in other ways, e.g. through exchanging scalar fields. For example, consider the possible 3-loop diagram in Fig.~\ref{fig:kineticmix3loop}. Because both particles and their antiparticles run in this loop, the diagram will vanish: the loop on the right with $\psi^+$ will be canceled by a loop with $\psi^-$.

%%%%%%%%%%%%%%%%%%%
%%%%%%%%%%%%%%%%%%%

%%%%%%%%%%%%%%%%%%%
\begin{figure}[!h]
\centering
\includegraphics[width=0.8\textwidth]{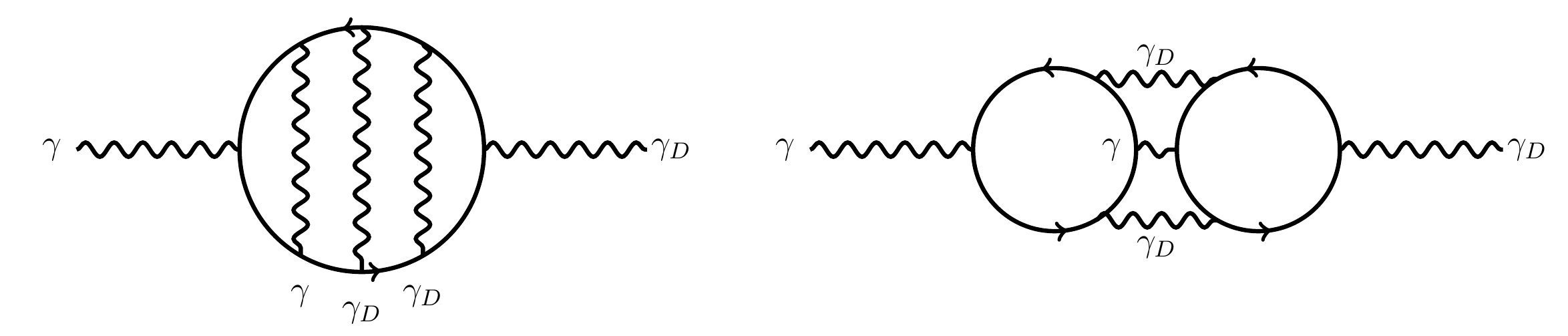}
\caption{Possible 4-loop contributions to kinetic mixing when there are particles charged under both U(1)s. At left: this diagram is proportional to ${\rm Tr}(Q^3 Q_D^5)$. The condition that ${\rm Tr}_{Q_D}(Q^3) = 0$, i.e. that the trace of visible charged cubed vanishes in the sector with any given dark charge, is sufficient to make this diagram vanish. At right: this diagram is proportional to ${\rm Tr}(Q^2 Q_D^2){\rm Tr}(Q Q_D^3)$. The condition that ${\rm Tr}_Q(Q_D^3) = 0$ is sufficient for it to be zero.}
\label{fig:kineticmix4loop}
\end{figure}
%%%%%%%%%%%%%%%%%%%

In fact, kinetic mixing can vanish to high loop order even if there {\em are} particles charged under the two groups, by choosing charge assignments to satisfy certain anomaly-like conditions. The one-loop diagram is proportional to ${\rm Tr}(Q Q_D)$, with $Q$ the visible charge of the particle running in the loop and $Q_D$ the dark charge. If {\em for every choice of $Q$} the sum of the dark charge of particles with visible charge $Q$ vanishes, a condition that we denote ${\rm Tr}_Q(Q_D) = 0$, the one-loop kinetic mixing will vanish. At higher loops there are diagrams with extra photon and dark photon exchanges going as ${\rm Tr}(Q^m Q_D^n)$ for $m, n \geq 1$. An example of such a diagram at four loops is shown at left in Figure~\ref{fig:kineticmix4loop}. If we impose the conditions ${\rm Tr}_Q(Q_D) = {\rm Tr}_Q(Q_D^3) = {\rm Tr}_{Q_D}(Q) = {\rm Tr}_{Q_D}(Q^3) = 0$, all of these diagrams with four or fewer loops will vanish. Another set of diagrams at four loops involves two fermion loops, like the right-hand plot in Fig.~\ref{fig:kineticmix4loop}, but these are set to zero by the same trace constraints. These are anomaly-like constraints in the sense that they demand certain vanishing traces, but they are much more restrictive than anomalies: they apply to scalars in the loop as well as to fermions, and they restrict the trace of dark charges in the sector with fixed visible charge and vice versa. Although these conditions are restrictive, they could potentially be satisfied, and could forbid kinetic mixing up to 5 loops. Such models would be consistent with the bound in Eq.~\ref{eq:kappabound}.

Another possibility is that U(1)$_Y$ or U(1)$_D$ is embedded in a nonabelian group. For example, suppose that U(1)$_D$ arises from a group SU(2)$_D$ broken by an adjoint Higgs $\Phi_D$. If the lightest particle charged under both SU(2)$_D$ and U(1)$_Y$ has mass $M > \left<\Phi_D\right>$, we expect that the mixing arises from a dark $S$-parameter operator~\cite{ArkaniHamed:2008qp}
\beq
\frac{g_D g'}{16\pi^2} \frac{1}{M} {\rm Tr}(\Phi^a_D W^a_{D\mu\nu}) B^{\mu \nu}.
\eeq
This is consistent with the bound $\kappa \simlt 10^{-9}$ if, for example, SU(2)$_D$ is broken at the weak scale and all particles charged under both groups have masses above $10^9$ GeV. In such a scenario, the threat that GUT- or string-scale physics renders the model inconsistent with the bounds can be avoided.

Another distinctive scenario is to consider that an {\em unbroken} nonabelian dark force remains at low energies. In this case kinetic mixing with the photon is completely impossible. One might worry that such a force would confine and prevent long-range interactions. However, with the relatively small values of $\alpha_D$ at which cooling can be effective, the temperature of our dark plasma in galaxies will be too high for confinement to occur. Nonabelian sectors that don't confine because they flow to infrared fixed points are also a possibility, but in this case we would need more fields with dark charge and the BBN bounds on the number of light degrees of freedom would become more severe.

One final possibility is that the U(1)$_D$ is not exact and the gauge boson is not massless.

\subsection{Other Interactions Between $C,{\bar C}$ and the SM Plasma}

\begin{figure}[h]
\centering
\includegraphics[width=0.75\textwidth]{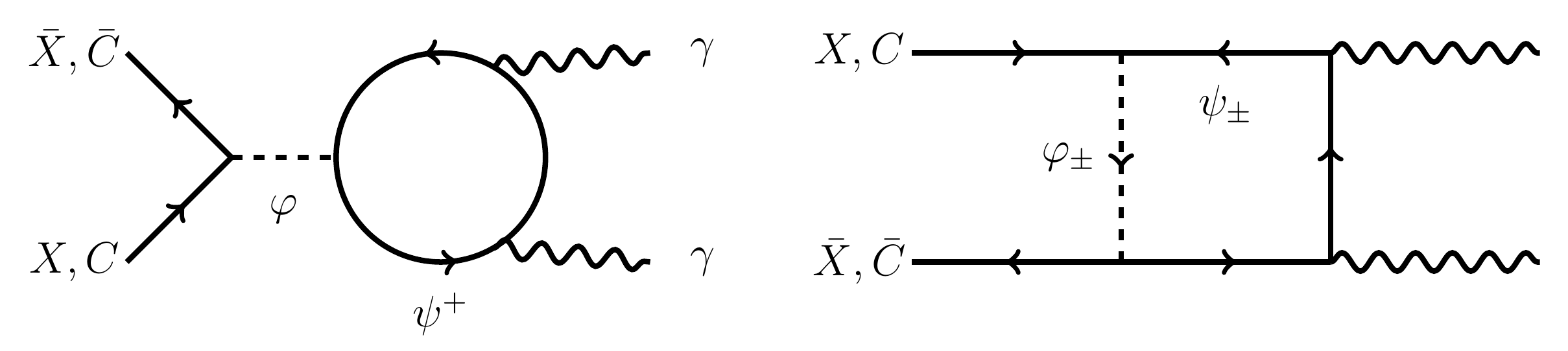}
\caption{Possible scenarios for fermionic dark matter giving rise to an annihilation signal in gamma rays. At left: $s$-channel intermediate scalar. (It could also be a spin-one $Z'$ with a $\gamma Z$ final state). At right: box topology with charged intermediate states.}
\label{fig:fermionannihilationmodes}
\end{figure}

In models with indirect detection signals, we assume that the heavy dark sector particles $X$ and ${\bar X}$ can annihilate to Standard Model particles; for instance, we can consider the process $X{\bar X} \to \gamma\gamma$. If the light particles $C$ have couplings similar to those of $X$, then the inverse process $\gamma\gamma \to C{\bar C}$ can produce relativistic $C$ particles at late times. This may be in conflict with the bound on light degrees of freedom discussed in Sec.~\ref{subsec:temp}. Unlike kinetic mixing, this constraint would arise from a higher dimension operator,
\beq
\frac{1}{\Lambda^3} C{\bar C}F_{\mu\nu}F^{\mu\nu},
\eeq
which is dimensionally suppressed and imposes weaker constaints than in the previous subsection. The cross section is
\beq
\sigma(\gamma \gamma \to C{\bar C}) \approx \frac{T^4}{4\pi \Lambda^6}.
\eeq
This scattering process potentially keeps $C$ and ${\bar C}$ in equilibrium with the Standard Model. To check this, we compare to the Hubble rate:
\beq
\frac{n_\gamma \sigma(\gamma \gamma \to C{\bar C})}{H} \approx \frac{\frac{2\zeta(3)}{\pi^2} T^3 \frac{T^4}{4\pi \Lambda^6}}{\sqrt{\frac{\pi^2}{90} g_*} \frac{T^2}{M_{\rm Pl}}} \approx 0.06 g_*^{-1/2} \frac{T^5 M_{\rm Pl}}{\Lambda^6}.
\eeq
This shows that this process no longer couples $C$ to the Standard Model thermal plasma when the temperature drops below
\beq
T \approx \left(\frac{g_*^{1/2} \Lambda^6}{0.06 M_{\rm Pl}}\right)^{1/5} \sim 250~{\rm MeV} \left(\frac{\Lambda}{200~{\rm GeV}}\right)^{6/5} \left(\frac{g_*}{100}\right)^{1/10}.
\eeq
Thus, if the coupling of $X$ to the dark sector is fixed through mediators with mass of order $\Lambda$ at the weak scale, $\gamma\gamma \to C{\bar C}$ scattering is approximately consistent with the hidden sector decoupling from the visible sector at temperatures above 200 MeV, which we found to be safe in Section~\ref{sec:history}. Many models for the UV completion of this annihilation model are possible, including those illustrated in Fig.~\ref{fig:fermionannihilationmodes}. In fact, for a fermionic model chiral symmetry can be used to further lower the couplings of $C$ relative to those of $X$, rendering the model even safer from constraints from $\gamma\gamma \to C{\bar C}$.

\begin{figure}[h]
\centering
\includegraphics[width=0.4\textwidth]{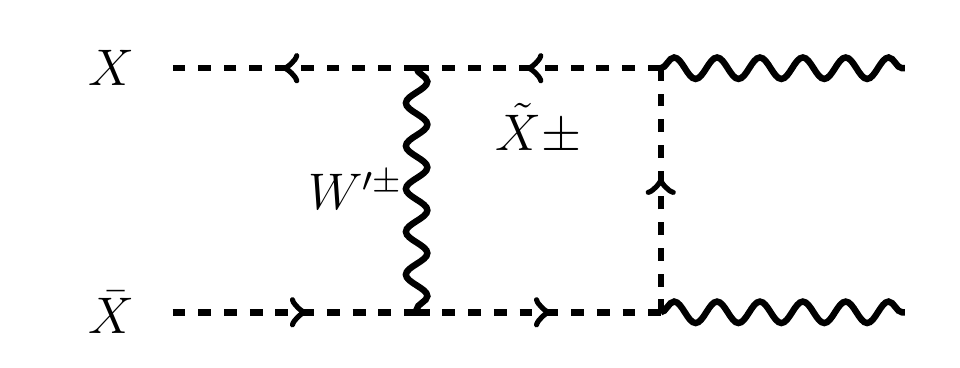}
\caption{Possible scenario for an annihilation signal in gamma rays: scalar loop diagram with a box topology. This can give $X$ annihilation and not $C$ production, consistent with bounds, if $C$ does not couple to the new gauge boson $W'$.}
\label{fig:boxtopology}
\end{figure}

Notice that for scalar dark matter $\phi$, the operator $\phi^\dagger \phi F_{\mu \nu} F^{\mu \nu}$ has dimension six and $\phi$ would not decouple from the Standard Model until lower temperatures, posing a potential problem for $\Delta N^{\rm BBN}_{{\rm eff},\nu}$. These models however could be safe as well if the UV completion works as in Fig~\ref{fig:boxtopology}, if the coupling to a new gauge boson $W'$ is absent for the $C$ particles (so that they are not produced from the Standard Model plasma) but not for the $X$ particles (so that they can annihilate to photons).

%%%%%%%%%%%%%%%%%%%
%%%%%%%%%%%%%%%%%%%

%%%%%%%%%%%%%%%%%%%
%%%%%%%%%%%%%%%%%%%

\section*{Acknowledgments}
We would especially like to thank Lars Hernquist for guidance early on in this project. We also thank Adam Brown, Clifford Cheung, Roland de Putter, Daniel Eisenstein, Doug Finkbeiner, Liam Fitzpatrick, Josh Frieman, Shy Genel, Lawrence Hall, Jared Kaplan, Manoj Kaplinghat, John March-Russell, Philip Mauskopf, Matt McQuinn, Moti Milgrom, Ann Nelson, Yasunori Nomura, Adi Nusser, Josh Ruderman, Matt Schwartz, Tanmay Vachaspati, Matt Walker, and Rogier Windhorst for useful discussions. We thank Howard Georgi for supplying our title. We are supported in part by the Fundamental Laws Initiative of the Harvard Center for the Fundamental Laws of Nature. AK and MR thank the Galileo Galilei Institute for Theoretical Physics in Florence, Italy, for its hospitality while a portion of this work was completed. The work of LR was supported in part by NSF grants PHY-0855591 and PHY-1216270.


\begin{thebibliography}{999}
\bibitem{SteinhardtNewPaper}  We also recently learned of other work in progress with a separate interacting component without dissipation and with gravothermal collapse from Pollack and Steinhardt.
\end{thebibliography}
\end{document}